\documentstyle[epsfig]{mn}
\pagerange{\pageref{firstpage}--\pageref{lastpage}} \pubyear{2001}

\begin{document}
\twocolumn


\title{The role of multipolar magnetic fields in pulsar
  magnetospheres}
\author[Asseo \& Khechinashvili]{Estelle
  Asseo$^{1}$ and David Khechinashvili$^{1,2,3}$\\
  $^{1}$ Centre de Physique Th{\'e}orique, Ecole Polytechnique, 91128
  Palaiseau CEDEX, France\\
  $^{2}$ Institute of Astronomy, University of Zielona G{\'o}ra,
  Lubuska 2, 65-265 Zielona G{\'o}ra, Poland\\
  $^{3}$ Center of Plasma Astrophysics, Abastumani Astrophysical
  Observatory, Al. Kazbegi Ave. 2a, 380060 Tbilisi, Georgia\\}
\date{Received in original form}
\maketitle
\begin{abstract}
We explore the role of complex multipolar magnetic fields in
determining physical processes near the surface of rotation
powered pulsars. We model the actual magnetic field as the sum of
global dipolar and star-centered multipolar fields. In
configurations involving axially symmetric and uniform multipolar
fields, 'neutral points' and 'neutral lines' exist close to the
stellar surface. Also, the curvature radii of magnetic field lines
near the stellar surface can never be smaller than the stellar
radius, even for very high order multipoles. Consequently, such
configurations are unable to provide an efficient pair creation
process above pulsar polar caps, necessary for plasma mechanisms
of generation of pulsar radiation. In configurations involving
axially symmetric and non-uniform multipoles, the periphery of the
pulsar polar cap becomes fragmented into symmetrically distributed
narrow sub-regions where curvature radii of complex magnetic field
lines are less than the radius of the star. The pair production
process is only possible just above these 'favourable'
sub-regions. As a result, the pair plasma flow is confined within
narrow filaments regularly distributed around the margin of the
open magnetic flux tube. Such a magnetic topology allows us to
model the system of twenty isolated sub-beams observed in PSR
B$0943+10$ by Deshpande \& Rankin (1999, 2001). We suggest a
physical mechanism for the generation of pulsar radio emission in
the ensemble of finite sub-beams, based on specific instabilities.
We propose an explanation for the subpulse drift phenomenon
observed in some long-period pulsars.
\end{abstract}

\begin{keywords}
pulsars: magnetic fields. plasmas: instabilities, waves.
\end{keywords}
\section{Introduction}\label{Intro}
Observations and theory suggest that complex multipolar magnetic
fields prevail near the surface of neutron stars, and play an
important role in the physics of rotation powered pulsars.

The complexity of surface magnetic fields is determined by the
evolution of magnetic fields in neutron stars. According to one
scenario (e.g., Blandford, Applegate \& Hernquist 1983), the magnetic
field is generated by currents flowing in the thin outer crust of the
star, with thickness $\Delta r\ll R_{\mathrm s}$, where $R_{\mathrm
  s}\approx 10^6$~cm is the neutron star radius. In this case, the
superposition of magnetic multipoles of order $l\sim R_{\mathrm
  s}/\Delta r\geq 10$ dominates the exterior field (Arons 1993).
Assuming that the evolution of magnetic fields in isolated neutron
stars is due to ohmic decay, Mitra, Konar \& Bhattacharya (1999)
show that the evolution of such high-order multipolar components
($l\leq 25$) is very similar to that of the dipolar field: the
structure of the pulsar surface field does not vary much as the
pulsar ages.  An alternative approach by Ruderman (1991a,b,c)
attributes the surface magnetic field evolution to changes in the
magnetic field present in the core. Due to stresses, caused by the
spinning down or the spinning up of the neutron star, strong
enough to drive the crustal lattice beyond its yield strength,
moving magnetized crustal platelets are created and account for
'sunspot-like' clumps of magnetic field at the stellar surface.

One first reason for introducing non-dipolar magnetic fields close to
the stellar surface lies in the inability for the pure surface dipolar
field to provide sufficient pair production, as required for the
generation of pulsar emission. In effect, in their classical paper,
Ruderman \& Sutherland (1975) state that the value of the radius of
curvature of magnetic field lines, $\rho$, should be about $R_{\mathrm
  s} \approx 10^6$~cm near the stellar surface, in order to fulfill
the conditions for a copious pair production process in the polar
vacuum gap. However, such small curvature radii cannot be achieved
assuming a pure dipolar magnetic field near the stellar surface:
the contribution of non-dipolar magnetic components, present in
the close vicinity of the neutron star in the form of either
multipoles of order higher than the dipole one, or the
'sunspot-like' clumps, is necessary. These components are supposed
to be strong enough to change the local topology of the magnetic
field in the vicinity of pulsar polar caps.  Arons \& Scharlemann
(1979) also point out that pair creation in pulsars with very long
periods is impossible without invoking magnetic fields more
complex than the dipolar one. In a thorough study of the
space-charge limited flow of electrons in combined surface dipolar
and quadrupolar magnetic fields, Barnard \& Arons (1982) conclude
that the curvature radii of resulting magnetic field lines can be
much smaller than those of the pure dipolar field, and discuss the
possible impact of this effect on both the pair creation process
and the observations.

Chen \& Ruderman (1993) specify the conditions for a sufficient
pair production to be triggered off, as required for the
generation of pulsar emission. They conclude that only a very
twisted ($\rho_{\mathrm s}\sim R_{\mathrm s}$, where
$\rho_{\mathrm s}$ is the curvature radius at the stellar surface)
and strong ($B_{\mathrm s} = 2\times 10^{13}$~G) magnetic field
ensures that on the $P - {\dot P}$ diagram, the death line leaves
all pulsars above itself, that is to say, 'permits' their
existence. However, an exception is the recently discovered PSR
J$2144-3933$ with period $P=8.5$~s (Young, Manchester \& Johnston
1999), which appears to lie even below this death line. With this
latter observation in mind, Gil \& Mitra (2001) study anew the
pair creation process in a super-strong magnetic field. They
conclude that all observed pulsars, including PSR J$2144-3933$,
lie above the death lines deduced from their model, which adopts
$B_{\mathrm s} / B_{\mathrm s}^{\mathrm d} \sim 100$ (although
still $B_{\mathrm s} < B_{\mathrm q}$) and $\rho_{\mathrm s}\sim
0.1 R_{\mathrm s}$.  Here, $B_{\mathrm q} \equiv m^2 c^3/ e \hbar
= 4.4 \times 10^{13}$~G is the critical magnetic field strength
above which the photon splitting phenomenon may inhibit the pair
formation process (Daugherty \& Harding 1983), while $B_{\mathrm
s}^{\mathrm d} $ and $ B_{\mathrm s}$ are respectively the dipolar
and the total -  dipolar plus multipolar - magnetic field
strengths at the stellar surface. Thus, according to Gil \& Mitra
(2001), such extremely strong and curved magnetic fields are able
to resolve in a natural way the long-standing controversy about
the so-called 'binding energy problem', from which vacuum gap
models suffered so far, without the necessity to involve bare
polar cap strange stars, as proposed by Xu, Qiao \& Zhang (1999).

Unfortunately, only the dipolar component of the surface magnetic
field, more exactly, only its component perpendicular to the line
of sight, is inferred from the observed spin-down rate of a
pulsar, thus leaving the possibility for different hypotheses. For
instance, Krolik (1991) assumes that the rotational energy loss of
a pulsar $I\Omega{\dot\Omega}$ results from the electromagnetic
radiation generated by the ensemble of individual magnetic
multipoles which are dominant at the pulsar light cylinder: this
allows the surface value of each multipolar component to be
constrained, so yielding limits on magnetic moments of high order;
the tightest limits are obtained for millisecond pulsars, due to
their short periods $P$ and the small values of their derivatives
$\dot P$.  Such restrictions on high-order field strengths at the
light cylinder do not necessarily mean their absence close to the
stellar surface, where they can even play a dominant role. Krolik
(1991) suggests that, if such is the case, the relative complexity
of the profiles of millisecond pulsars results from the complexity
of their actual magnetic field.

Recent interpretations of the observations of soft X-rays, probably
emitted due to the heating of pulsar polar caps, lead to similar
conclusions. Page \& Sarmiento (1995, 1996) for instance, show that it
is impossible to obtain the high level of observed pulsed fractions,
or to describe asymmetric X-ray profiles, under the assumption of a
purely dipolar field: a much better fit to the observations is
obtained by including a suitably oriented quadrupolar field component.
On the other hand, Cheng \& Zhang (1999) in their model suggest that a
strong and complex surface magnetic field significantly affects the
two hard thermal components, usually present in the X-ray emission of
pulsars.

Interest in a multipole field dominant close to the star goes back
to  Gil (1985), who introduced a dominant quadrupolar magnetic field
in order to explain peculiarities in the radio emission of the
millisecond PSR $1937+214$.

Recently observed features of radio emission of the pulsar PSR
B$0943+10$, reported by Deshpande \& Rankin (1999, 2001), clearly
indicate the existence of a system of 20 rotating sub-beams, in
circulation around the magnetic axis of the star. This well
organized and stable pattern of emission has been observed at
three different frequencies, namely at 430~MHz, 111.5~MHz (at
Arecibo) and 34~MHz (in India), which are supposed to represent
three different altitudes above the stellar surface, according to
the Radhakrishnan \& Cooke (1969) model. This system of emission
sub-beams can be associated with a set of plasma columns, with
their feet in the acceleration region. The relativistic pair
plasma flow in these columns is guided by the strong dipolar
magnetic field up to (and beyond) the altitudes corresponding to
the radio emission generation region. At these altitudes,
instabilities occurring in a particular plasma column give rise to
a radio emission beam, independently of similar processes in
adjacent columns. These beams form the emission cone. Deshpande \&
Rankin (2001) point out that the columns themselves are not
drifting or rotating: it is the action in the acceleration region,
located near the stellar surface, which is doing so while it feeds
and maintains the plasma column just above it.  Several processes
in the individual sub-beams are found to be stable: circulation
time, intensities, polarisation characteristics and dimensions.
However, as noticed by Deshpande \& Rankin (2001), there is no
clear choice between a continuous drift effect and particle
cascade sputters. They speculate that the memory of the sub-beam
configuration may reside in either standing waves, or in heating
of the surface (varying along the magnetic azimuth), or in the
seeding of the plasma columns, for instance, by sparks, created on
a non-integer multiple of the circumference and thus apparently
drifting.

One way of describing successfully the observed set of discrete
emission columns in the pulsar magnetosphere is to assume that
close to the stellar surface the total magnetic field results from
the superposition of a dipolar and a high-order multipolar
components, both fields being axially symmetric and star-centered.
Moreover, to fulfill the requirements for an efficient pair
creation just above the surface of the star, namely the
possibility to have at the stellar surface a very twisted total
magnetic field, one may consider that the strength of the assumed
multipolar magnetic field is higher than the dipolar field
strength, or at least comparable to it, and test that the radius
of curvature of resulting magnetic field lines is less than the
stellar radius. This concerns the domain of open field lines, as
the latter includes the region where the emission process is
supposed to arise (at large distances from the stellar centre,
$r\gg R_{\mathrm s}$). In this same domain, near the stellar
surface, high magnetic multipolar components are supposed to be
more efficient than the dipolar field in determining the
properties of the vacuum gap and initiating the pair production
process. Let us remark that by `vacuum gap` is meant the domain
located just above the surface of the star in which a non-zero
component of the electric field in the direction parallel to the
background magnetic field is able to accelerate charges.

The likelihood of having a multipolar magnetic field with strength
higher than, or comparable to the dipole field strength is
reasonable if one takes into account the influence of the
general-relativistic effect of inertial frame dragging on the
electromagnetic field structure near a rotating neutron star, as
suggested by Muslimov \& Tsygan (1986). In Section~\ref{Topology:
intensification} of this paper we estimate such amplification
factors, using the formalism developed by Muslimov \& Tsygan
(1986), and confirm that these factors increase with the
multipolar order $l$ of the multipolar components involved.
Therefore, the actual strength of the high-order multipolar field
components near the stellar surface may appear to be much higher
than those estimated in the flat space-time.  On the other hand,
as mentioned above, the evolution of multipolar magnetic fields
with $l\leq 25$ due to ohmic diffusion follows the same pattern as
that of the dipolar field (Mitra, Konar \& Bhattacharya 1999).
Therefore, there are reasonable grounds to believe that the
dipolar and multipolar magnetic field components coexist and
persist with their relative strengths.

The above configuration can be used to explore pulsar radio
emission theories. For instance, the pulsar emission cone, when
associated with a stable configuration of an ensemble of emission
columns, could possibly be traced back to spark discharges in the
pulsar polar gap region. This could be done in reference to recent
works involving sparks and non-stationary plasma in vacuum gap
models (Usov 1987; Ursov \& Usov 1988; Asseo \& Melikidze 1998;
Melikidze, Gil \& Pataraya 2000; Gil \& Mitra 2001). In a
different way, we model the whole system of observed twenty
sub-beams as an ensemble of relativistic finite beams, flowing
along the curved magnetic field lines of the open magnetosphere
and immersed in external media. The two surrounding external media
which limit these flowing sub-beams are, on one side, the medium
that lies on closed field lines, and on the other side, the
tenuous plasma present inside the hollow cone. The latter
surrounds the pulsar magnetic axis, and is modified due to the
presence of high-order multipolar magnetic field components near
the stellar surface.

We show how, in the case of millisecond or fast pulsars, the
geometrical characteristics of the flowing beams in the radial and
azimuthal directions allow us to treat each beam of the ensemble
as isolated and immersed in external media.  The complex
dispersion relation for perturbations of the system is simplified,
as the beams are extremely thin and as the propagating waves are
excited close to resonance in the radio domain. Such waves have
simultaneously an electrostatic and electromagnetic character and
thus are able to reach an observer. Depending on the distance of
the considered region to the surface of the star, either the
'finite beam', or the radiative, or the two-stream instability is
the dominant one (Asseo 1995). Moreover, whatever the altitude and
the concerned instability, unstable waves with characteristics in
agreement with observed waves can be considered as candidates for
pulsar radio emission. In particular, close to the surface of the
star, above the gap region in which the observed beams are
supposed to have their origin, the 'finite beam' instability,
dominant there, may initiate the radio emission process in the
plasma columns. Farther away the classical emission processes are
available.

In the case of slower pulsars, like PSR $0943+10$, the geometrical
characteristics of the flowing beams in the radial and azimuthal
directions prevent considering each particular beam of the
ensemble as isolated, just above the gap region and up to some
distance $r_{_{\mathrm 2D}}$.  There, all the beams forming the
emitting region have to be treated simultaneously.  This can be
done by introducing the 3-dimensional dependence of the perturbed
physical quantities, leading to a simple analytical treatment of
the whole system of sub-beams. In between the gap region and the
region located at and above $r_{_{\mathrm 2D}}$, simultaneous
dependence of the perturbations on the radial and azimuthal
variables will complicate the character of the emitted waves. In
effect, the 3-dimensional character of perturbed physical
quantities results in the coupling of the features of excited
waves in the radial and azimuthal directions. Beyond the distance
$r_{_{\mathrm 2D}}$, the sub-beams of the ensemble can be
considered as isolated, so that the resulting perturbations should
resemble those described above for fast pulsars: the character of
the unstable emitted waves should be recovered.  However, the
existence of the necessary transition between the two regions,
below and above $r_{_{\mathrm 2D}}$, and thus the required
continuity of the wave-solutions through this transition zone,
suggest the possibility for an apparent 'drift' of the observed
waves.

In the present paper we examine several magnetic configurations
involving dipolar and strong multipolar magnetic field components,
in order to determine those which provide small curvature radii,
as required by standard magnetospheric models. Note that our
calculations only concern the simplified case of the so-called
aligned rotator, i.e. a pulsar whose dipolar magnetic momentum
$\bmu$ is parallel, or anti-parallel, to the rotation axis
${\bmath \Omega}$, and similarly for the magnetic momenta of the
different multipoles. However, we believe that most of our
qualitative results also apply to inclined rotators.

As a starting point we suppose that magnetic multipoles of order
$(l,m)$, created by the currents flowing in the stellar crust
(Blandford et~al. 1983; Arons 1993), are superimposed on the total
dipolar field. A multipole component with $m = 0$ we refer to as
'axially symmetric and uniform', and the one with a non-zero $m$
-- as 'axially symmetric and non-uniform'. Let us note that
uniformity, and not the symmetry, is the main difference between
these two structures of the magnetic field near the polar cap.
Indeed, while $m=0$ -type fields are simultaneously symmetric and
uniform, $m \neq 0$ -type fields are symmetric around the magnetic
axis but non-uniform, being fragmented as shown below. In
Section~\ref{Fields}, we present the general formalism to describe
the multipolar fields. We also consider the configurations
composed of the global dipolar field and a single axially
symmetric and uniform multipolar field, and find out that 'neutral
points' and 'neutral lines', i.e. places of zero magnetic field
strength, may exist near the neutron star surface in such a
geometry. In Section~\ref{Topology}, we model the magnetic
structure of complex configurations in the vicinity of the stellar
surface, and calculate the curvature radii of resulting field
lines near the stellar surface. We conclude that the curvature
radii of field lines in configurations containing only axially
symmetric and uniform multipolar fields are not small enough to
ignite the pair creation process above the pulsar polar cap, even
if contributing multipoles are of very high order. On the other
hand, we find that configurations involving axially symmetric and
non-uniform multipolar magnetic fields allow to achieve the small
curvature radii of field lines, within a fraction of the modified
polar cap. Furthermore, the pair plasma flow at large altitudes
from the stellar surface appears to be fragmented into a system of
isolated filaments, distributed symmetrically around the pulsar
magnetic axis, around the margin of the open dipolar flux tube. We
apply this model to PSR~B$0943+10$ and find resemblance with the
observed pattern of its radio emission (Deshpande \& Rankin 1999,
2001). In Section~\ref{Model}, we discuss radio emission processes
in the system of discrete sub-beams of dense plasma, surrounded by
media of reduced density, in relation with an analysis previously
proposed by Asseo (1995). This is done for either millisecond and
fast pulsars or for slower pulsars like PSR B$0943+10$.  A summary
of our results is presented in Section~\ref{Summary}.

\section{Multipolar magnetic field components around the neutron
  star}\label{Fields}
Throughout this paper we use a system of spherical coordinates ($r,
\theta, \phi$) where $r$ is the distance measured from the stellar
centre, $\theta$ is the polar angle (in radians) measured from the
$z$-axis, and $\phi$ is the azimuthal angle (in radians) measured from
an arbitrary origin. The $z$-axis is directed along the dipolar
momentum of the star $\bmu$.

\subsection{General description of multipolar fields at close
  distances from the stellar surface}
\label{Fields: general}

In the near zone limit ($kr\ll 1$), the magnetic field vector,
${\bmath B}^{lm} (r,\theta,\phi)$, associated with a given magnetic
multipole $(l m)$, can be expressed in terms of spherical harmonics
$Y_{l m}(\theta,\phi)$ (Jackson 1975):
\begin{equation}
{\bmath B}^{lm}(r,\theta,\phi) =
\nabla\left(\frac{Y_{lm} (\theta,\phi)}
{r^{l+1}}\right) .
\label{B_lm}
\end{equation}
Spherical harmonics are written in terms of the associated Legendre
functions, as
\begin{equation}
Y_{lm}\left(\theta,\phi \right) = \sqrt{\frac{2l+1}{4\pi}
\frac{\left(l-m\right) !}{\left(l+m\right) !}}
P^m_l\left(\cos\theta\right) {\mathrm e}^{{\mathrm i}m\phi},
\label{Y_lm}
\end{equation}
where $P^m_l\left( x \right)$, the associated Legendre function, is
defined as
\begin{equation}
P^m_l\left( x \right) =\frac{\left( -1 \right)^m}{2^l l~!} \left(
1-x^2\right)^{m/2}\frac{{\mathrm d}^{l+m}}{{\mathrm d}
x^{l+m}}\left(x^2-1\right)^l
\label{P_lm}.
\end{equation}

In a spherical geometry, the components of an individual multipolar
magnetic field vector ${\bmath B}^{lm}(r,\theta,\phi) = (B_{
  r}^{lm}(r,\theta,\phi), B_{\theta}^{lm}(r,\theta,\phi),
B_{\phi}^{lm}(r,\theta,\phi))$ are written:

\begin{equation}
B_{ r}^{lm}(r,\theta,\phi)
 = - 4 \pi  \frac {l+1}{2 l+1}  \frac{ q_{lm}}{r^{l+2}}
Y_{lm}(\theta,\phi),
\label{B_r}
\end{equation}
\begin{equation}
B_{\theta}^{lm}(r,\theta,\phi)
=  \frac {4 \pi}{ 2 l+1}  \frac{ q_{lm}}{r^{l+2}}
Y_{lm}^{(1,0)}(\theta,\phi),
\label{B_theta}
\end{equation}
\begin{equation}
B_{\phi}^{lm}(r,\theta,\phi)
=   \frac {4 \pi}{2 l+1}  \frac{q_{lm}}{ r^{l+2}}
 {\mathrm i} m \sin{\theta}  Y_{lm}(\theta,\phi).
\label{B_phi}
\end{equation}

Particular multipolar magnetic components are obtained for
specific values of $l$ and $m$, with $|m|\leq l$. Namely, dipolar
components are obtained for $l=1$ and either $m=0$ or $m=\pm 1$,
quadrupolar components for $l=2$ and either $m=0$, $m= \pm 1$ or
$m= \pm 2 $ and so on. If $m=0$ and $l$ is arbitrary, then one
obtains axially symmetric and  uniform multipolar components.

The magnetic moment, associated with the strength of the magnetic
multipole component $(l m)$ at the surface of the star, is defined as
$\mu^{lm} = B_{\mathrm s}^{lm} R_{\mathrm s}^{l+2}$.

The total magnetic field of the neutron star can be written as a sum
of the individual multipolar magnetic field components. In the
hypothesis where all the multipoles are centered at the centre of the
star,
\begin{equation}
{\bmath B}(r,\theta,\phi) =
\sum_{l=1}^{\infty} \sum_{m=0,\pm 1,\ldots\pm l}
{\bmath B}^{lm}(r,\theta,\phi).
\label{B_total}
\end{equation}

Magnetic field lines for the total magnetic field are obtained, as
usual, from an integration of the set of differential equations:
\begin{equation}
\frac{{\mathrm d} r}{B_{r}(r,\theta,\phi)} = \frac { r {\mathrm
d}\theta }{B_{\theta}(r,\theta,\phi)} = \frac{r \sin{\theta}
{\mathrm d}\phi }{B_{\phi}(r,\theta,\phi)}. \label{equations}
\end{equation}

A unit vector in the direction of the local magnetic field ${\bmath
  B} (r,\theta,\phi)$ can be defined as:
\begin{equation}
{\bmath b}(r,\theta,\phi) \equiv
\frac{{\bmath B}(r,\theta,\phi)}{B (r,\theta,\phi)}.
\label{b}
\end{equation}

The absolute value of the derivative of ${\bmath b} (r,\theta,\phi)$
along the field line is nothing else than the curvature of the
magnetic field line. At a given point of space $(r,\theta,\phi)$, the
radius of curvature of the total magnetic field line is therefore
defined as
\begin{equation}
\rho (r,\theta,\phi)=\frac{1}{|\left({\bmath b}(r,\theta,\phi)
\nabla \right){\bmath b}(r,\theta,\phi)|}.
\label{rho}
\end{equation}

Naturally, if one assumes the existence of only one pure
individual multipolar magnetic field component around the star,
the equations of its field lines and their curvature radii can be
easily obtained from equations~(\ref{equations}) and (\ref{rho})
by substituting ${\bmath B}^{lm}(r,\theta,\phi)$ instead of the
total field ${\bmath B}(r,\theta,\phi)$.

\subsection{The case of
axially symmetric and uniform
multipolar magnetic fields:
magnetic field components and the magnetic vector potential}
\label{Fields: m=0}

In this section we treat multipolar magnetic fields of type ${\bmath
  B}_{l0}$, i.e. those for which $l$ is arbitrary but $m=0$. Such
fields can be regarded to as 'dipolar-like', as they are
axially symmetric and uniform
in the vicinity of pulsar polar caps, like the dipolar
magnetic field.


According to equations (\ref{B_r} $-$ \ref{B_total}), for the case
with $m=0$, the components of the total magnetic field close to the
surface of the star can be expressed in a general form as a sum of
multipolar magnetic field components, namely
\begin{eqnarray}
&&B_{r}(r,\theta) = \sum_{l= 1}^{\infty} B_{r}^{l
  0}(r,\theta)= \nonumber \\
&&\!\!\!\sum_{l= 1}^{\infty}\frac{\mu ^{l0}}{r^{l+2}}
\ f^{l 0}\!\!\left(\sum_{n = 0}^{2 n \leq l} c_{l - 2 n}
\cos^{(l- 2 n)}{\theta}\right),
\label{B_r m=0}
\end{eqnarray}
\begin{eqnarray}
&&B_{\theta}(r,\theta) =
\sum_{l= 1}^{\infty} B_{\theta}^{l 0}(r,\theta)=\nonumber \\
&&\!\!\!\sum_{l= 1}^{\infty} \frac{ c^{l 0} \mu^{l 0} \sin{\theta}}{r^{l+2}}
\ F^{l 0}\!\!\left(\sum_{n = 0}^{2 n + 1 \leq l} c_{l -1 - 2 n}
\cos^{(l - 1 - 2 n)}{\theta}\right),
\label{B_theta m=0}
\end{eqnarray}
\begin{eqnarray}
&&B_{\phi}(r,\theta) =  \sum_{l= 1}^{\infty} B_{\phi}^{l 0}(r,\theta)
= 0,
\label{B_phi m=0}
\end{eqnarray}
where $\mu^{l 0} = B_{\mathrm s}^{l 0} R_{\mathrm s}^{l+2}$ is the
magnetic moment associated with $B_{\mathrm s}^{l 0}$, the
strength of the multipolar magnetic field component at the surface
of the star, $\mu^{l 0}$ being a positive quantity according to
its definition; $c^{l 0}$ is a constant which depends on the order
$l$ of the multipole; $f^{l 0}$ and $F^{l 0}$ are specific
functions which only depend on powers of $\cos{\theta}$; $c_{l - 2
n}$ and $c_{l -1 - 2 n}$ are numerical coefficients.

As implied by Maxwell's equations, the static magnetic field
vector can be defined through the magnetic vector potential
$\bmath A$, as ${\bmath B}=\left(\nabla {\mathbf \times} {\bmath
A}\right)$. Thus, one can associate a vector potential ${\bmath
A^{l m}}$ with any arbitrary magnetic multipole with indices $l$
and $m$.  Obviously, in the axially symmetric and uniform case
with arbitrary $l$ and with $m = 0$, each multipolar component of
the magnetic field vector, ${\bmath B}^{l 0}(r,\theta)$, derives
from a potential vector ${\bmath A}^{l 0}(r,\theta)$. Assuming
that the magnetic field is created by an extremely thin current
that circulates all over the crust of the neutron star, the
potential vector has only an azimuthal component
$A_{\phi}(r,\theta)$, which can be specified for each particular
magnetic multipole having indices $l\geq 1$ and $m = 0$, in terms
of the corresponding current, or equivalently in terms of the
magnetic moment $\mu^{l 0}$ and the variables $\theta$ and $r$, as
\begin{equation}
A_{\phi}^{l 0}(r,\theta) =  \frac{\mu^{l 0} \sin
{\theta}}{r^{l+1}} \ F^{l 0}\!\!\left(\sum_{n = 0}^{2 n + 1 \leq
l} c_{l -1 - 2 n} \cos^{(l - 1 - 2 n)}{\theta}\right),
\label{Aphil0}
\end{equation}
where the functions $F^{l 0}$ have been introduced just above.

Magnetic field lines are obtained as usual from an integration of
the set of two differential equations (\ref{equations}). Here, for
the axially symmetric and uniform case with arbitrary $l$ and $m =
0$, the $\phi$-component of the magnetic field being zero, it is
sufficient to consider the multipole magnetic field components in
a plane $\phi={\mathrm const}$, and to integrate only the first
differential equation, namely
\begin{equation}
\frac{{\mathrm d}r}{B_{r}^{l 0}(r,\theta )}
= \frac {r {\mathrm d}\theta }{B_{\theta}^{l 0}(r,\theta )}.
\label{equations phi=0}
\end{equation}

A systematic derivation of the equations of field lines of any
particular magnetic multipole can also be obtained, introducing the
function
\begin{equation}
 \Psi^{l 0}(r,\theta) \equiv r \sin{\theta} A_{\phi}^{l 0} (r,\theta),
\label{Psi}
\end{equation}
which is an integral of the field equations. Effectively, using the
fact that magnetic field components of the multipolar magnetic field
with indices ($l 0$) derive from a vector potential $A_{\phi}^{l 0}
(r,\theta)$, we obtain:
\begin{equation}
B_{r}^{l 0}(r,\theta)
=  \frac{1}{r \sin{\theta}}  \frac{\partial}{\partial \theta}
\left(A_{\phi}^{l 0}(r,\theta) \sin{\theta}\right) =
 \frac{1}{r^{2} \sin{\theta}}  \frac{\partial}{\partial \theta}
\Psi^{l 0}(r,\theta),
\label{B_r_F}
\end{equation}
\begin{equation}
B_{\theta}^{l 0}(r,\theta) =  -\frac{1}{r}  \frac{\partial}{\partial r}
\left( r A_{\phi}^{l 0}(r,\theta)\right) =
 -\frac{1}{r \sin{\theta}}  \frac{\partial}{\partial r}
\Psi^{l 0}(r,\theta),
\label{B_theta_F}
\end{equation}
\begin{equation}
B_{\phi}^{l 0}(r,\theta) = 0.
\label{B_phi_F}
\end{equation}

The function $\Psi^{l 0}(r,\theta)$ is constant along a given
field line; the behaviour of multipolar magnetic field components
$B_{r}^{l 0}(r,\theta)$ and $B_{\theta}^{l 0}(r,\theta)$,
necessarily follows the behaviour of the derivatives
$\partial_r\Psi^{l 0}(r,\theta)$ and $\partial_{\theta}\Psi^{l
0}(r,\theta)$ along a given field line and in a given plane in
which $\phi={\mathrm const}$. As a consequence,
\begin{equation}
\Psi^{l 0}(r,\theta) = {\rm const}
\label{Psi=const_}
\end{equation}
gives the general form for multipolar magnetic field line
equations. For a magnetic multipole $(l 0)$, from
equations~(\ref{Aphil0}) and (\ref{Psi}),
equation~(\ref{Psi=const_}) can be written
\begin{equation}
\frac{\mu^{l 0} \sin^{2}{\theta}}{r^l}
\ F^{l 0}\!\!\left(\sum_{n = 0}^{2 n + 1 \leq l} c_{l -1 - 2 n}
\cos^{(l - 1 - 2 n)}{\theta}\right) = {\mathrm const}.
\label{Psi=const}
\end{equation}

The functions $F^{l 0}$ are very easily obtained for the successive
multipolar components: $F^{1 0 } = 1$, $F^{2 0 } = \cos{\theta}$,
$F^{3 0 } = 3 + 5 \cos{2 \theta}$, \dots, $F^{l 0} = \sum_{n = 0}^{2 n
  + 1 \leq l} c_{l -1 - 2 n} \cos{(l - 1 - 2 n)}{\theta}$, making use
of the transformations of circular functions $F^{l 0}\!\!\left(\sum_{n
    = 0}^{2 n + 1 \leq l} c_{l -1 - 2 n} \cos^{(l - 1 - 2
    n)}{\theta}\right) \equiv F^{l 0}\!\!\left(\sum_{n = 0}^{2 n + 1
    \leq l} c'_{l -1 - 2 n} \cos {(l - 1 - 2 n) \theta}\right)$. These
functions, according to equation (\ref{Psi=const}), provide magnetic
field line equations for the appropriate particular multipolar fields.
Drawing of the field lines in a plane where $\phi= {\mathrm const}$,
shows that a pure magnetic multipole of order $l$ makes $2 l$ loops
around the centre of the star.

A function $\Psi (r, \theta)$ can be associated with the total
magnetic field: considering that the total magnetic field, in a
given plane $\phi= {\mathrm const}$, is the sum of the successive
multipolar magnetic field components, the function $\Psi (r,
\theta)$ is the sum of the successive functions $\Psi^{l 0}(r,
\theta) $ defined above. The general form of the function $\Psi
(r, \theta)$ can be written
\begin{equation}
\Psi(r,\theta)  = \sum_{l = 1}^{\infty} \Psi^{l 0}(r,\theta)=
 r \sin{\theta} \sum_{l = 1}^{\infty} A_{\phi}^{l 0}(r,\theta).
\end{equation}
Consequently, the total function $\Psi(r,\theta)$, written as an
infinite sum,
\begin{equation}
\Psi(r,\theta) = \sin^2{\theta}
\sum_{l= 1}^{\infty}  \frac{\mu^{l 0}}{r^l}
\ F^{l 0}\!\!\left(\sum_{n = 0}^{2 n + 1 \leq l} c_{l -1 - 2 n}
\cos^{(l - 1 - 2 n)}{\theta}\right),
\end{equation}
provides the total magnetic field line equation $\Psi(r, \theta) =
{\rm const}$ in the given plane $\phi= {\mathrm const}$.

Using the specific contributions of the first multipoles, together
with the functions $F^{l 0}$ introduced in equation~(\ref{B_theta
  m=0}) (see also the text below equation~\ref{Psi=const}), and with
the adequate numerical coefficients, we have:
\begin{eqnarray}
\Psi(r,\theta) = \sin^2{\theta} \left[ \frac{\mu^{1 0 }}{r}
 + \frac{\mu^{2 0}}{r^2} \cos{\theta}
  +  \frac{\mu^{3 0}}{r^3} (3 + 5 \cos{2 \theta})
  + \dots \right. \nonumber \\
\left. + \frac{\mu^{l 0}}{r^l}\ F^{l 0}
\!\!\left(\sum_{n = 0}^{2 n + 1 \leq l}
c'_{l -1 - 2 n} \cos{(l - 1 - 2 n)}{\theta}\right)+ \ldots \right].
\end{eqnarray}

\subsection{The case of
axially symmetric and uniform
multipolar magnetic fields:
neutral lines, neutral points}\label{Fields: neutral points}

The surface $\Psi (r, \theta) = 0 $ separates field lines with
opposite direction, that is field lines on which the magnetic field
has reversals. Whatever the number of multipoles, the function
$\Psi(r,\theta)$ is zero for either $\theta = 0$, or $\theta = \pi$,
or when the variables $r$ and $\theta$ are such that $\sum_{l =
  1}^{\infty} Y_{l 0}^{(1,0)}(\theta,\phi) = 0$. Obviously, there
should be some particular relation between the magnetic moments
$\mu^{l 0}$ to have $\Psi(r,\theta) = 0$, and thus a total
magnetic field with zero strength, $B(r,\theta)=0$, in any
arbitrary plane $\phi = {\mathrm const}$ and at some point $(r,
\theta)$, as, for instance, at the surface of the star, where $r =
R_{\mathrm s}$.

Values of the magnetic field on the lines $\theta = 0$, or $\theta
= \pi$ are easily obtained. From the equations~(\ref{B_r m=0}) --
(\ref{B_phi m=0}), giving the magnetic field components, it is
clear that on the lines $\theta = 0$ and $\theta = \pi$,
$B_{\phi}(r,\theta)= 0$, as it is zero everywhere,
$B_{\theta}(r,\theta) = 0$, since $\sin{\theta}$ is in factor, but
$B_{r}(r,\theta)$ takes-on values that depend on the particular
point where the radial component of the multipolar magnetic field
is calculated. In particular, on the line $\theta = \pi$, there is
a point at which $B_{r}(r,\theta)= 0$. A study of the behaviour of
the magnetic field around this point will probably show that it
may be considered as a neutral point for particular configurations
of the magnetic field in the pulsar magnetosphere.  For instance,
this should be the case when the total magnetic field can be
represented by a combination of either dipolar and quadrupolar
magnetic fields, or dipolar and octupolar magnetic field, or
dipolar and 16-polar magnetic field and so on, the equations for
the field lines being, respectively,
\begin{equation}
\Psi (r, \theta) = \sin^2{\theta} \left(\frac{\mu^{1 0}}{r}
+ \frac{\mu^{2 0}}{r^2} \cos{\theta}\right),
\end{equation}
\begin{equation}
\Psi (r, \theta) = \sin^2{\theta} \left(\frac{\mu^{1 0}}{r}
+\frac{\mu^{3 0}}{r^3} (3 + 5 \cos{2 \theta}) \right),
\end{equation}
\begin{equation}
\Psi (r, \theta) = \sin^2{\theta} \left(\frac{\mu^{1 0}}{r}
+ \frac{\mu^{4 0}}{r^4}(9 \cos{\theta}+ 7 \cos{3 \theta}) \right),
\end{equation}
and so on.

On the side of the pole where $\theta = 0$, there is no real
solution for $r$, the $r$-solutions being either negative or
imaginary. By contrast, on the side of the other pole, where
$\theta = \pi$, there is a point $r$ at which the magnetic field
strength is zero, and changes sign around this point. Effectively,
for the combination of a dipole plus a multipole of even order $(l
= 2 n, n = 1, 2,\ldots)$, there is a point where the strength of
the magnetic field is zero. The solution $r_{\mathrm np}^{l1} =
{\mathrm (numerical\ coefficient)} \left(B_{\mathrm s}^{l0}/
B_{\mathrm s}^{10}\right)^{1/(l-1)} R_{\mathrm s},$  depends on
the ratio of the multipolar and the dipolar magnetic field
strengths. More generally, for an arbitrary $\theta$-angle there
is a neutral line on which the magnetic field is zero. For
instance, if the magnetic field is taken as the simple combination
of dipolar and quadrupolar magnetic fields, the function
\begin{equation}
\Psi (r, \theta) =   \sin^2 {\theta}
\left( \frac{\mu^{1 0}}{r}  + \frac{\mu^{2 0}}{r^2} \cos{\theta}\right)
\end{equation}
becomes zero at either $ \theta = 0 $ or $ \theta = \pi$,
or on the line where
\begin{equation}
\frac{\mu^{1 0}}{r}  + \frac{\mu^{2 0}}{r^2} \cos{\theta} = 0,
\end{equation}
that is to say, on the circle defined by the equation
\begin{equation}
\cos{\theta}= - r \frac {\mu^{1 0}} {\mu^{2 0}}
= - \frac{r}{R_{\mathrm s}}
\frac{ B_{\mathrm s}^{10} }{ B_{\mathrm s}^{20}}.
\end{equation}

In this case, there are three neutral lines: the two axes, $\theta =
0$ and $\theta = \pi$ and a circle in the plane $(r, \theta)$, on
which the only acceptable $r$-solutions correspond to $\theta$-angles
in the domain $[\pi/2, 3 \pi/2]$.  Indeed, depending on the respective
values of $\mu_{1 0}$ and $\mu_{2 0}$, the $r$-solutions on the
neutral circle have a physical meaning, or none.
As an example, for the combination of dipolar and quadrupolar
magnetic fields, the 'neutral point' is located at distances from
the centre of the star $r_{\mathrm np}^{21}= 1.16 \left(B_{\mathrm
s}^{20}/B_{\mathrm s}^{10}\right)R_{\mathrm s}$. On this line
$\theta=\pi$, $B_{r}$ is positive at smaller distances, $r <
r_{\mathrm np}^{21}$, and negative at larger distances, $r >
r_{\mathrm np}^{21}$. The 'neutral point' is located at the
surface of the star only if $B_{\mathrm s}^{20}$ is less than
$B_{\mathrm s}^{10}$, so that $B_{\mathrm s}^{20} = B_{\mathrm
s}^{10}/1.16 = 0.86 B_{\mathrm s}^{10}$.

As clear from the general results given above, similar locations for
the 'neutral points' can be easily obtained for any combination of a
dipolar field plus a multipolar field with an even index $l$.
A study of the behaviour of the magnetic field on lines close to the
'neutral line' is necessary to conclude about the specific physical
processes in this region.

\section{Modelling the magnetic field structure near the stellar
  surface: topology of possible configurations and curvature radii of
  field lines}\label{Topology}

\subsection{General description of the method of calculation}
\label{Topology: general} In this section we study the topology of
magnetic field configurations near the stellar surface, which
consist of a multipolar magnetic field ${\bmath B}^{lm}$ (with
$l>1$ and $|m|\leq l$), superimposed on the global dipolar field
of the pulsar ${\bmath B}^{\mathrm d}\equiv {\bmath B}^{10}$. We
also introduce the ratio of the total and the dipolar magnetic
field strengths,
\begin{equation}
  \label{beta}
\beta(r,\theta,\phi)\equiv \frac{B(r,\theta,\phi)}
{B^{\mathrm d}(r,\theta,\phi)},
\end{equation}
where $B (r,\theta,\phi) \equiv | {\bmath B}^{\mathrm d}
(r,\theta,\phi) + {\bmath B}^{lm}(r,\theta,\phi)| $. Therefore,
$\beta(r,\theta,\phi)$ indicates how many times the strength of
the total field exceeds the strength of its dipolar component, at
a given point of space $(r,\theta,\phi)$. In the calculations
reported in Sections~\ref{Topology: m=0} and \ref{Topology: m/=0},
particular attention is paid to the cases where the curvature
radius of magnetic field lines at the stellar surface is smaller
than the stellar radius, i.e. $\rho_{\mathrm s} \leq R_{\mathrm
s}$, at least within a fraction of the modified polar cap.
Intuitively, one would expect that such configurations assume a
significant contribution of the multipolar component, so that
$\beta_{\mathrm s} \equiv \beta(R_{\mathrm
  s},\theta,\phi) > 1$.

In order to test various magnetic configurations, we redefine the
expression for the multipolar field components (equation~\ref{B_lm})
by introducing an extra amplification factor $\xi$ as a free
parameter, so that this expression now takes the following form:
\begin{equation}
{\bmath B}_{lm}(r,\theta,\phi) = \xi \nabla\left(\frac{Y_{lm}
(\theta,\phi)} {r^{l+1}}\right) .
\label{B_lm_new}
\end{equation}
A specific choice of the $\xi$-factor in every particular example
considered below, is justified by the output: it should allow us to
obtain the expected structure and strength of the surface magnetic
field above the modified polar cap. Amplification of certain
multipolar components is discussed in detail in Section~\ref{Topology:
intensification}.

In order to model the geometry of the total magnetic field, one
has to solve the set of differential equations~(\ref{equations}),
which is now convenient to rewrite in the following form:
\begin{eqnarray}
&& r \frac{{\mathrm d} \theta (r)}{{\mathrm d} r}  =  \frac{B_\theta
  \left( r,\theta(r), \phi (r)\right)}{B_r \left( r,\theta (r),\phi (r)\right)}
\label{field lines1},\\
&& r\sin\theta (r)\frac{{\mathrm d} \phi(r)}{{\mathrm d} r}  =  \frac{B_\phi
\left(r,\theta (r),\phi (r)\right)} {B_r \left( r,\theta (r),\phi (r)\right)}.
\label{field lines2}
\end{eqnarray}
This set of equations has to be solved for the unknown functions
$\theta(r)$ and $\phi(r)$.

Of course, the complex surface magnetic field should transform
into the global pure dipolar field at large altitudes from the
stellar surface, where multipolar components effectively vanish.
This is warranted by the choice of correct boundary conditions
while solving the set of equations~(\ref{field lines1}) $-$
(\ref{field lines2}). Since the generation of radio emission is
associated with the bundle of the open dipolar magnetic field
lines (i.e. those crossing the light cylinder), we restrict our
calculations to this domain of the magnetosphere.

Let us introduce some distance $r_{_{\mathrm i}}$, where surface
fields are negligibly small, $B^{lm}(r_{_{\mathrm i}},\theta
(r_{_{\mathrm i}}),\phi (r_{_{\mathrm i}}))\ll B^{\mathrm d}
(r_{_{\mathrm i}},\theta (r_{_{\mathrm i}}),\phi (r_{_{\mathrm
    i}}))$, so that the total magnetic field can be treated as purely
dipolar there.  At this distance, the points of the last dipolar open
field lines delineate a certain circle, and their azimuthal angle
$\phi (r_{_{\mathrm i}})$ takes
arbitrary values within the range
$0\leq\phi (r_{_{\mathrm i}})\leq 2\pi$. According to the equation for
the dipolar field lines,
\begin{equation}
\frac{r}{\sin^{2}{\theta}} = {\rm const},
\end{equation}
the polar angle of these points can be written
\begin{equation}
  \label{theta_max}
\theta_{_{\mathrm i}} =\arcsin\sqrt{\frac{r_{_{\mathrm i}}}
{R_{_{\mathrm LC}}}}\approx \arcsin \left(1.45\times
10^{-2}{P^{-0.5}} \sqrt{\frac{r_{_{\mathrm i}}}{R_{\mathrm
s}}}\right).
\end{equation}
Here $R_{_{\mathrm LC}}= Pc/2\pi$ is the light cylinder radius of the
pulsar, and $P$ is the pulsar period. The dipolar field falls off with
the distance as $\left( R_{\mathrm s}/r\right)^3$, whereas the
multipolar field of order $l$ falls off as $\left( R_{\mathrm
    s}/r\right)^{l+2}$, according to equations~(\ref{B_r}) $-$
(\ref{B_phi}). Therefore, using equation~(\ref{beta}), the
characteristic distance $r_{_{\mathrm B}}$, where $B^{lm}\sim
B^{\mathrm d}$, can roughly be estimated as
\begin{equation}
r_{_{\mathrm B}} \sim (\beta_{\mathrm s} - 1)^{1/(l-1)}
R_{\mathrm s}.
\label{r_B}
\end{equation}
Thus, the dipolar field dominates over the multipolar field in the
domain of altitudes $r\gg r_{_{\mathrm B}}$ (and vice versa).
Choice of $r_{_{\mathrm i}}$ in the latter domain warrants a
smooth transition from the complex surface field to the open
dipolar field lines. Hence, this is equivalent to the integration
of the field line equations from the stellar surface up to the
light cylinder.

As well-known, observations of pulsar radio emission and most of the
models for its generation imply that it originates at altitudes $r\geq
50 R_{\mathrm s}$. Interpretation of radio observations indicates that
the magnetic field is purely dipolar in the generation region, so that
complex surface fields should vanish at such
altitudes\footnote{Although, the anomalous dispersion measure of the
  high-frequency versus low-frequency radio pulses, observed in some
  pulsars (Davies et~al.  1984; Kuzmin 1992), has been attributed to
  the presence of the quadrupolar component of the magnetic field in
  the emission region.}. In other words, the complex surface field
lines should connect to pure dipolar field lines well below $r=50
R_{\mathrm s}$. We verify that this requirement is fulfilled
through out our calculations, for each chosen $\xi$-parameter.

Hence, the boundary conditions at $r = r_{_{\mathrm i}}$ for different
field lines are to be taken within the following angular range:
\begin{eqnarray}
  \label{init_cond}
&&0\leq \theta (r_{_{\mathrm i}})\leq \theta_{_{\mathrm i}}, \nonumber\\
&&0\leq \phi (r_{_{\mathrm i}})\leq 2 \pi.
\end{eqnarray}
Each set of boundary conditions defines a single open field line,
starting at the stellar surface, where the multipolar magnetic
field dominates over the dipolar field, and extending to the
domain located at altitudes $r \geq r_{_{\mathrm i}}$ where the
field is purely dipolar and axially symmetric and uniform. We use
a Fehlberg fourth-fifth order Runge-Kutta method (see, e. g.,
Forsythe et al. 1977) in order to obtain the numerical solutions
of the set of equations~(\ref{field lines1}) -- (\ref{field
lines2}).

The results of our simulations are presented in the following
subsections. We separate them into two cases, which have diverse
physical consequences, namely: i) configurations composed of
axially symmetric and uniform multipolar magnetic fields with
$m=0$ and the global dipolar field, and ii) configurations
composed of axially symmetric and non-uniform multipolar fields
with $m\neq 0$ and the global dipolar field. Before starting to
discuss these particular cases, let us present one of the
arguments in favour of an anomalous amplification of the 'weight'
of multipolar components at the stellar surface, with respect to
the dipole field.

\subsection{Intensification of multipolar magnetic field components due
  to the general relativistic effect of inertial frame
  dragging}\label{Topology: intensification}

As mentioned in Section~\ref{Intro}, high-order multipolar fields
may be intensified close to the surface of a rotating neutron
star, relatively to their value in a flat space, due to the
general-relativistic effect of inertial frame dragging studied by
Muslimov \& Tsygan (1986). The exterior gravitational field of a
neutron star, rotating with a constant angular velocity $\Omega$,
is described by the metric, which is locally inertial at the
infinity:
\begin{equation}
{\mathrm d}s^2 = g_{\alpha \beta} {\mathrm d}x^{\alpha} {\mathrm
  d}x^{\beta},\ \ \alpha,\beta=0,1,2,3.
\end{equation}
The nonzero components of the metric tensor are written as
follows:
\begin{eqnarray}
g_{0 0} = -g_{11}=  1 - \frac{R_{\mathrm g}}{r}, \
g_{2 2}= - r^2, \nonumber \\
g_{3 3}= - r^2  \sin^2 {\theta}, \
g_{0 3}= \frac{2 G  J }{c^3 r} \sin^2 {\theta}.
\label{metric tensor}
\end{eqnarray}
Here $R_{\mathrm g}= 2 G M/c^2$ is the Schwarzschild radius, $G$ is
the gravitational constant, $c$ is the speed of light, $M$ is the mass
of the neutron star and $J$ is its angular momentum.
\begin{figure}
  \unitlength1.0cm
  \begin{center}
    \begin{picture}(17.,8.3)
      \put(-0.1,-0.6){
        \epsfysize=9cm
        \epsffile{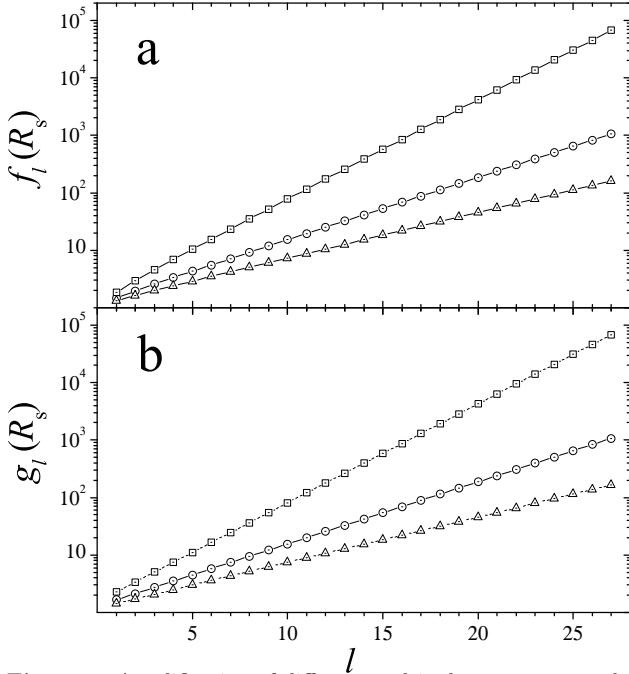}}
    \end{picture}
  \end{center}
\caption{Amplification of different multipolar components due to
the frame dragging effect (equations~\ref{B_r grav} and
\ref{B_theta grav}). The values of the functions $f_{l}(r)$ and
$\sqrt{g_{00}}g_{l}(r)$ at the stellar surface ($r=R_{\mathrm
s}$), given by equations~(\ref{f_l}) and (\ref{g_l}), are plotted
versus $l$ in panels (a) and (b), respectively. This plot
corresponds to the neutron star with the presumed mass $M=1.4
M_{\odot}$. The curves with squares, circles and triangles
represent the calculations for the presumed stellar radii of
$7\times 10^5$~cm, $10^6$~cm and $1.3\times 10^6$~cm,
respectively.} \label{f_g_l}
\end{figure}
\begin{figure}
  \unitlength1.0cm
  \begin{center}
    \begin{picture}(17.,8.3)
      \put(-0.1,-0.6){
        \epsfysize=9cm
        \epsffile{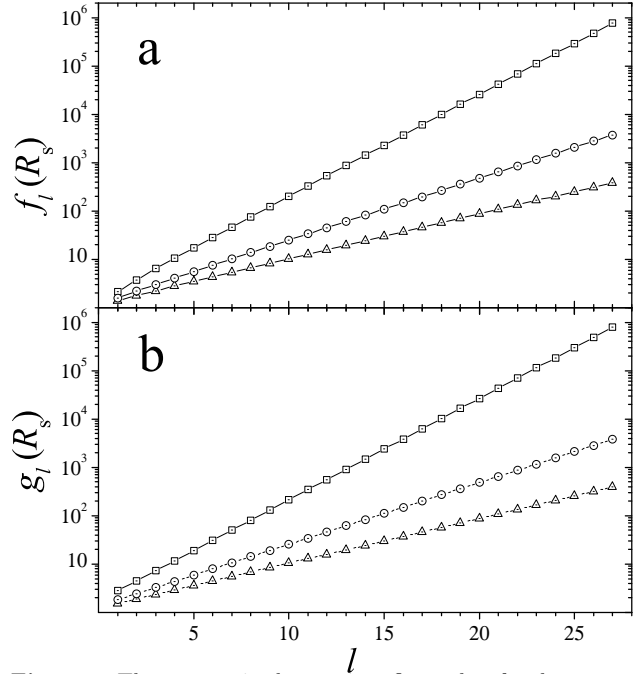}}
    \end{picture}
  \end{center}
\caption{The same as in the previous figure, but for the neutron
    star with the presumed mass $M=1.6 M_{\odot}$.} \label{f_g_l1}
\end{figure}

Muslimov \& Tsygan (1986) solved the Maxwell equations in such a
geometry, and found the modified magnetic field components modified as
follows:
\begin{equation}
{\tilde B}_r(r,\theta,\phi) = \sum_{lm} (l+1)
\left(\frac{R_{\mathrm s}}{r}\right)^{l+2} \!\!f_l(r)\  b_{lm}\,
Y_{lm}(\theta,\lambda),
\end{equation}
\begin{equation}
{\tilde B}_{\theta,\phi}(r,\theta,\phi) = \sum_{lm}
\left(\frac{R_{\mathrm s}}{r}\right)^{l+2}
\!\!\sqrt{g_{00}} g_l(r)\ b_{lm}\,
\nabla_{\theta,\phi}Y_{lm}(\theta,\lambda).
\end{equation}
Here $b_{lm}$ are some coefficients defined from the boundary
conditions (see equations~\ref{B_r} -- \ref{B_phi}),
$\lambda=\phi-\Omega t$, and the functions $f_{l}(r)$ and $g_{l}(r)$
can be expressed in terms of hyper-geometric functions,
\begin{equation}
f_{l}(r) = F\left(l,l+2;2(l+1);  \frac{R_{\mathrm g}}{r} \right),
\label{f_l}
\end{equation}
\begin{equation}
g_{l}(r) = F\left(l+1,l+2;2(l+1); \frac{R_{\mathrm g}}{r}\right).
\label{g_l}
\end{equation}

Therefore, the modified radial and angular multipolar magnetic field
components are connected with the corresponding components in the
flat space, as
\begin{equation}
{\tilde B}_{r}^{lm} = f_{l}(r) B_{r}^{lm}, \label{B_r grav}
\end{equation}
\begin{equation}
{\tilde B}_{\theta,\phi}^{lm} = \sqrt{g_{00}}g_{l}(r)  B_{\theta,\phi}^{lm}.
\label{B_theta grav}
\end{equation}

Estimates of the values of the functions $f_{l}(r) $ and
$\sqrt{g_{00}} g_{l}(r)$ at the stellar surface, $r=R_{\mathrm
s}$, are displayed in Figs.~\ref{f_g_l} and \ref{f_g_l1}, for a
presumed mass of the neutron star $M=1.4 M_{\odot}$ and $M=1.6
M_{\odot}$, respectively. The functions $f_{l}(r)$ and $g_{l}(r)$
are calculated for three different cases of the model neutron star
radii, namely $R_{\mathrm s}=7\times 10^5$~cm, $10^6$~cm and
$1.3\times 10^6$~cm. First, one notices that the resulting
intensification is about the same for the radial and angular
magnetic field components, and is stronger for the higher-order
multipoles. Second, multipolar fields are amplified more
intensively near the surface of relatively smaller and more
massive neutron stars. Indeed, according to Fig.~\ref{f_g_l1},
high-order multipolar magnetic field components with $l\leq 25$ in
the curved space may be even $10^4 - 10^5$ stronger than the
corresponding components in the flat space.

The evolution with time of multipolar magnetic field strengths in
isolated neutron stars has been studied by Mitra et~al. (1999),
assuming that the field is mostly generated in the outer crust of
the star and its later evolution is due to the ohmic decay of
currents in the crustal layers. They have shown that if one
assumes the same initial strength for all multipolar magnetic
fields, the evolution is such that the reduction with time is
similar to that of the dipole field component, except for very
high multipole orders ($l > 25$).

A combination of the two described effects, that is, the
intensification of multipolar magnetic field components due to
general relativistic effects and evolution with time of their
strengths, suggests that multipolar magnetic field components with
high order should be present at the surface of neutron stars and
important enough to determine the magnetic configuration in their
vicinity. This justifies the choice of $\xi$
(equation~\ref{B_lm_new}) as a free parameter in our calculations
below.

\subsection{Complex configurations containing
axially symmetric and uniform multipolar magnetic fields: field
line topology and curvature radii} \label{Topology: m=0} In this
section we present the results of our calculations for the
magnetic configurations composed of the global dipolar field and a
surface multipolar field with $m=0$ and arbitrary $l\geq 1$. As
mentioned in the beginning of Section~\ref{Fields: m=0}, such
fields can be regarded as 'dipolar-like', as they are axially
symmetric and uniform in the vicinity of pulsar polar caps. Due to
this fact, the ratio of the total and the dipolar magnetic field
strengths at the stellar surface, $\beta_{\mathrm s}$
(equation~\ref{beta}), is independent of the azimuth $\phi$ and
also almost independent of the polar angle $\theta$, within the
modified polar cap of a pulsar. It should be noticed that the
polar cap still has a circular shape, as in the pure dipolar case.
However, as a result of magnetic flux conservation, the presence
of a strong surface magnetic field with $\beta_{\mathrm s} > 1$
reduces its angular radius to $\theta_p^{l0}=\theta_p^{\mathrm d}
\beta_{\mathrm s}^{-0.5}$, compared to that of the pure dipolar
polar cap, $\theta_p^{\mathrm d}\approx 1.45\times 10^{-2}
P^{-0.5}$~[rad].

For pure multipolar magnetic field lines with $m=0$ (i.e., in the
absence of the dipolar component), a first-order approximation for the
curvature radius is obtained from equation~(\ref{rho}), as
\begin{equation}
  \label{rho_m=00}
\rho^{l0} \approx \frac{4}{l (l+2)}\frac{r}{\theta}.
\end{equation}
For example, the curvature radius of the last open field lines at the
stellar surface ($r=R_{\mathrm s}, \theta =\theta_{p}^{l0}$) for a
pure multipole $(l0)$ is:
\begin{equation}
 \label{rho_m=0}
\rho_{\mathrm s}^{l0} \approx  \frac{276\;\beta_{\mathrm s}^{0.5}
P^{0.5}}{l (l+2)} R_{\mathrm s}.
\end{equation}
However, as demonstrated below, the presence of the dipolar magnetic
field in the actual configuration near the stellar surface modifies
the values given by equation~(\ref{rho_m=0}), especially assuming
small and moderate values of $\beta_{\mathrm s}$.
\begin{figure}
  \unitlength1.0cm
  \begin{center}
    \begin{picture}(17.,12.3)
      \put(-0.1,-0.6){
        \epsfysize=13cm
        \epsffile{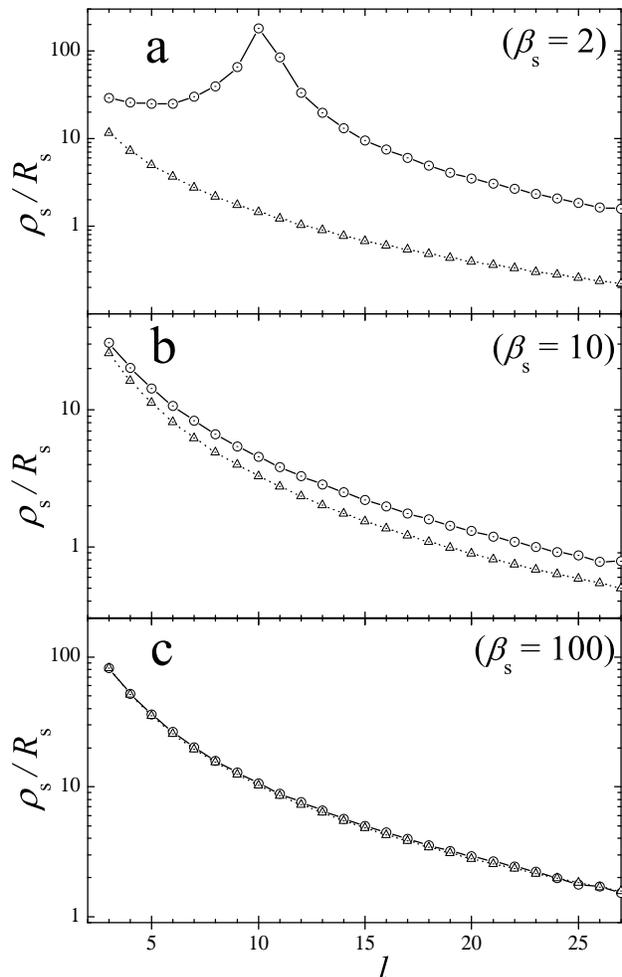}}
    \end{picture}
  \end{center}
\caption{Curvature radii of the last open field lines at the
stellar surface in the configurations composed of the global
dipolar field and the axially symmetric and uniform multipolar
field with $m=0$, as functions of the multipolar order $l$. The
three panels (a), (b) and (c) correspond to the three different
cases of relative contribution of the multipolar components to the
surface field strength, $\beta_{\mathrm s}=2$, $\beta_{\mathrm
s}=10$ and $\beta_{\mathrm s}=100$, respectively. The solid lines
with circles correspond to the calculations for the complex
surface field, whereas the dotted lines with triangles represent
curvature radii, as they would be in the absence of the dipolar
field (equation~\ref{rho_m=0}).}
 \label{l_R_c_all}
\end{figure}

As an example, we consider a sample pulsar with $P=0.2$~s and
$\dot P=10^{-15}$~s/s. The surface dipolar magnetic field of such
a pulsar is estimated as $B_{\mathrm s}^{\mathrm d}\approx
3.2\times 10^{19}\left(P \dot P\right)^{0.5}\approx 4.4\times
10^{11}$~G.  Such a value allows intensification of the surface
field up to $100$ times, without exceeding the critical field
$B_{\mathrm q}=4.4\times 10^{13}$~G. We model the dipole plus
single multipole magnetic field lines of this pulsar, according to
the scheme described in Section~\ref{Topology: general}, and
calculate the curvature radii of the last open field lines at the
stellar surface.  Our aim is to check how strong should be the
contribution of axially symmetric and uniform multipolar
components near the stellar surface, in order to achieve the
desired small values of curvature radii of the resulting open
magnetic field lines, as required for an efficient pair creation
process.

The curvature radii of the last open field lines at the stellar
surface are presented in Figs.~\ref{l_R_c_all}(a), (b) and (c), which
correspond to cases where the total surface magnetic field exceeds the
dipolar component 2, 10 and 100 times, respectively. The solid lines
with circles represent the results of our numerical calculations for
given complex configurations. The dotted lines with triangles show,
for comparison, what would be the curvature radii of the involved
multipolar magnetic field lines in the absence of the dipolar field
(as given by equation~\ref{rho_m=0}).

In Fig.~\ref{l_R_c_all}(a) it is seen that, if the contribution of
surface multipolar fields is comparable to the dipolar one,
$\beta_{\mathrm s}=2$, the radius of curvature of resulting field
lines remains larger than the stellar radius, $\rho_{\mathrm s} >
R_{\mathrm s}$, even for multipolar fields of very high order, $l
\geq 25$. On the other hand, the dotted line shows that, in the
absence of the dipolar field, $\rho_{\mathrm s} \leq R_{\mathrm
s}$ would be achieved already for multipoles with $l\geq 12$.
Therefore in this case, multipolar fields are simply too weak to
'twist' the field lines strongly enough.

The situation is somewhat better, if one assumes a stronger
contribution of multipolar fields, $\beta_{\mathrm s}=10$. This case
is presented in Fig.~\ref{l_R_c_all}(b). As seen in this figure, the
curvature radii of field lines become less than the stellar radius, if
the multipolar order of contributing multipoles is very high, $l\geq
24$, still quite a possible case, according to Mitra et~al. (1999).

However, a further increase of the contribution of surface
multipolar fields worsens the situation. For instance, in the case
of $\beta_{\mathrm s} = 100$, presented in
Fig.~\ref{l_R_c_all}(c), curvature radii never drop below the
stellar radius, even when very high-order multipolar components
($l \geq 25 $) contribute into the total magnetic field.  This
fact is easy to explain: such strong multipoles fully determine
the surface field structure near the stellar surface.  Indeed, the
dotted line in Fig.~\ref{l_R_c_all}(c), corresponding to the
absence of the dipolar component, almost follows the solid line,
corresponding to the total surface magnetic field. On the other
hand, the high surface field strength leads to a highly (10 times)
reduced angular radius of the polar cap, which, according to
equation~(\ref{rho_m=0}), provides larger curvature radii.

Therefore, we conclude that in the configurations composed of the
global dipolar field and the axially symmetric and uniform
multipolar fields, the curvature radii of the open field lines do
not generally have curvature radii, small enough to fulfill the
conditions for a copious pair creation process.
At the same time, as seen in Fig.~\ref{l_R_c_all}(c), in the
configurations composed of global dipole and axially symmetric and
uniform multipole, it is impossible to have both $\rho_{\mathrm s}
\leq R_{\mathrm s}$ and $\beta_{\mathrm s}\sim 100$ in the same
configuration, as implied by the vacuum gap model of Gil \& Mitra
(2001). These conclusions are true at least for multipolar fields
of order $l \leq 25$, which do not significantly dissipate during
the pulsar characteristic life-time (Mitra et~al. 1999).

\subsection{Complex configurations containing axially symmetric and non-uniform
multipolar magnetic fields: field line topology and curvature radii}
\label{Topology: m/=0}

Let us now study magnetic field configurations composed of a
global dipolar field on which non-axisymmetric multipolar magnetic
fields (i.e. those with $m\neq 0$) are superimposed. We again
restrict our calculations to the open magnetic flux tube. As in
the previous section, we start our calculations at some altitude
$r_{_{\mathrm i}}$ where $B^{lm}(r_{_{\mathrm i}},\theta,\phi)\ll
B^{\mathrm d} (r_{_{\mathrm i}} ,\theta,\phi)$, and solve the set
of equations~(\ref{field lines1}) and (\ref{field lines2}) from
$r_{_{\mathrm i}}$ down to the stellar surface. While doing so, we
take the boundary conditions in the range~(\ref{init_cond}). This
allows us to find the footprints of the tube of open field lines
on the stellar surface, i.e., to construct the modified polar cap.
Our aim is to get solutions which provide $\rho\leq R_{\mathrm s}$
at the stellar surface, at least within a fraction of the modified
polar cap, which is necessary for an efficient pair production
process. We adjust the free parameter $\xi$ (introduced in
equation~\ref{B_lm_new}) in each particular case considered below,
in order to obtain desired configurations.  Throughout our
calculations we control the value of the total surface magnetic
field, so that it does not exceed $B_{\mathrm q}$ within the polar
cap.  Obviously, there is a great variety of possible
configurations. Here we present the most illustrative ones. Let us
first make a few general remarks.

A pure multipolar magnetic field ($l,m$) with $m\neq 0$ has
$2m(l-m+1)$ poles over the whole stellar surface.  Out of those,
along each latitude containing poles, $2m$ poles of alternating
polarity are distributed around the axis of symmetry. At a given
latitude, each pole with a given polarity is neighboured on its
sides by two poles of the opposite polarity. The inclusion of the
global dipolar field changes this topology. In effect, the dipolar
field lines which enter one of the two polar regions on the
stellar surface, are only able to connect to the local multipolar
poles of a given polarity, and not to the poles of the opposite
polarity. This reduces to $m$ the number of local poles, which are
close to each of the polar regions. One expects that the influence
of these local multipolar poles is stronger at the annular edge of
the polar cap. Consequently, the polar cap becomes modified: its
edge becomes rather fragmented into $m$ independent 'hot spots',
as demonstrated below.

The most interesting cases, in our opinion, are those where the
maximum number of local poles surround the magnetic $z$-axis of
symmetry of the dipolar field at the closest angular distance, i.e.
those with $l=m$.  As an example, let us first consider the multipolar
magnetic field with $l=4$ and $m=4$, superimposed on the global
dipolar field.
\begin{figure}
  \unitlength1.0cm
  \begin{center}
    \begin{picture}(17.,5.0)
      \put(-0.1,-0.6){
        \epsfysize=5.5cm
        \epsffile{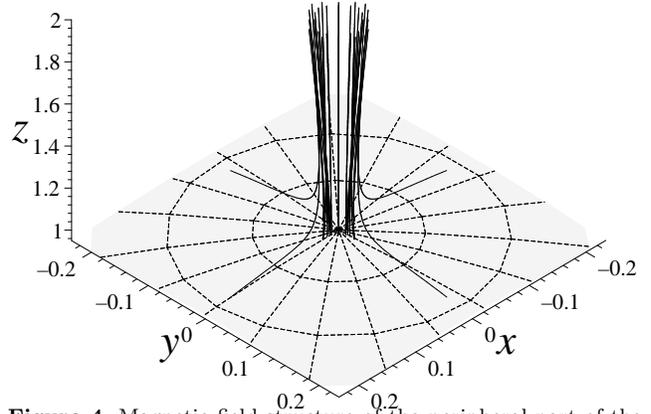}}
    \end{picture}
  \end{center}
\caption{Magnetic field structure of the peripheral part of
    the open flux tube near the stellar surface, corresponding to the
   mixture of the dipolar field and the multipole of order $l=4$ and
   $m=4$. The Cartesian coordinates $x$, $y$ and $z$ are represented in
   the units of the stellar radius $R_{\mathrm s}$.}
 \label{3dd}
\end{figure}
\begin{figure}
  \unitlength1.0cm
  \begin{center}
    \begin{picture}(17.,7.3)
      \put(-0.1,-0.6){
        \epsfysize=8cm
        \epsffile{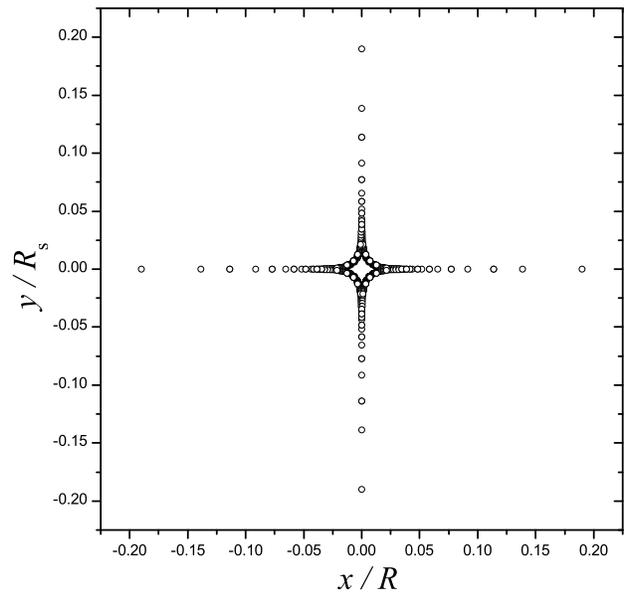}}
    \end{picture}
  \end{center}
\caption{Peripheral fraction of the modified polar cap for the
    contributing multipole $l=4,m=4$. The footprints of the open field
    lines are represented by circles. Deformation of the polar cap
    under the influence of the local multipolar poles is clearly
    seen.}
 \label{polcap_4}
\end{figure}

The open field line geometry for such a configuration near the
stellar surface and the corresponding modified polar cap (composed
of the footprints of the open field lines) are presented in
Figs.~\ref{3dd} and \ref{polcap_4}, respectively. The calculations
were done for a sample pulsar with $P=1$~s and $\dot
P=10^{-15}$~s/s. For simplicity, we only present in these figures
the outermost 0.2 fraction of the open flux tube and its
projection at the stellar surface (the inner, more central open
field lines are less affected by the multipolar field, and follow
rather the dipolar pattern). In other words, the boundary
conditions at an altitude $r_{_{\mathrm i}}=10 R_{\mathrm s}$ (see
equations~\ref{theta_max} and \ref{init_cond}) were taken within
the following angular ranges:
\begin{equation}
0\leq \phi(r_{_{\mathrm i}})\leq 2\pi,\ \ \ 0.8\theta_{_{\mathrm
i}}\leq \theta(r_{_{\mathrm i}})\leq \theta_{_{\mathrm i}}.
\label{init_cond_4}
\end{equation}
Each of these ranges was divided into a fixed number of equal parts,
both by the azimuth and the polar angle, so that the angular spacings
between the initial points at the altitude $r_{_{\mathrm i}}$ were
taken equal.

Fig.~\ref{3dd} shows that the bundle of last open field lines, which
is circumferentially uniform at higher altitudes, $r\geq 2 R_{\mathrm
  s}$, where the multipolar magnetic field is negligible, tends to
be deformed under the influence of the multipolar field closer to
the stellar surface. As it is also seen in Fig.~\ref{polcap_4},
the field lines are most significantly deviated outwards in those
$\phi$-directions, where the poles with an appropriate polarity
are located. On the contrary, in the directions towards the
multipolar poles of the opposite polarity, field lines deviate
inwards from the pure dipolar circular pattern. The resulting
polar cap appears to be rather distorted, as displayed in
Fig.~\ref{polcap_4}.

One would expect that both the magnetic field strength and the
curvature radius of the field lines also vary within the periphery
of the modified polar cap, so that both of them now appear to be
functions of $\theta$ and $\phi$, as opposed to the configurations
containing axially symmetric and uniform multipolar components
(discussed in Section~\ref{Topology: m=0}), where both
$\beta_{\mathrm s}$ and $\rho_{\mathrm s}$ could be treated as
constant throughout the modified polar cap. Let us study this
dependence.
\begin{figure}
  \unitlength1.0cm
  \begin{center}
    \begin{picture}(17.,12.3)
      \put(-0.1,-0.6){
        \epsfysize=13cm
        \epsffile{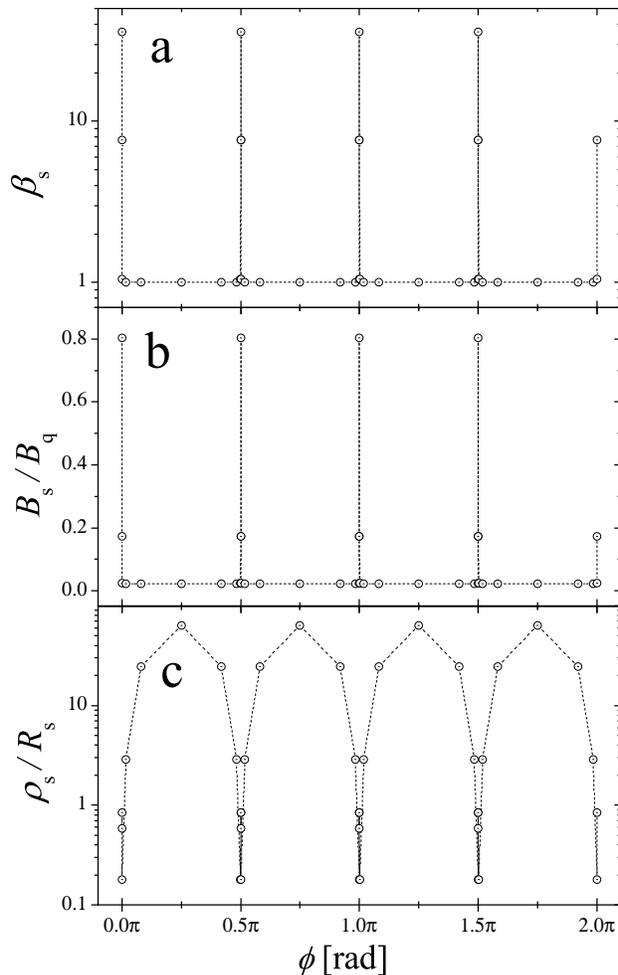}}
    \end{picture}
  \end{center}
\caption{Dependence of $\beta_{\mathrm s}$, $B_{\mathrm
    s}/B_{\mathrm q}$ and $\rho_{\mathrm s}$ on the azimuthal angle
    $\phi$, at the feet of the last open field lines, for the
   contributing multipole $l=4,m=4$. Conditions for pair creation are
   only fulfilled in localised regions, where
   $\rho_{\mathrm s} \leq R_{\mathrm s}$.}
 \label{all_phi_4}
\end{figure}

Fig.~\ref{all_phi_4}(a) displays the azimuthal variation of
$\beta_{\mathrm s}$ (i.e., the value of the total surface magnetic
field normalized to the local value of the dipolar component) at
the feet of the last open field lines. Fig.~\ref{all_phi_4}(b)
displays the same total magnetic field measured in the units of
the critical field $B_{\mathrm q}$. An apparent increase of the
magnetic field is seen in the direction of the appropriate
multipolar poles, whereas the field retains its almost dipolar
strength in between these peaks (the minimum field strength is
achieved towards the multipolar poles of the opposite polarity).
In Fig.~\ref{all_phi_4}(b) we see that the field may increase to a
significant fraction of $B_{\mathrm q}$ on certain field lines.

Fig.~\ref{all_phi_4}(c) shows the azimuthal variation of the
curvature radius (derived from equation~\ref{rho}) of the last
open field lines at the stellar surface. Clearly, $\rho_{\mathrm
s}\sim (0.1 \div 1)R_{\mathrm s}$ only for the group of field
lines stretched in the direction of the multipolar poles, whereas
the surface value of the curvature radius of the remaining field
lines (directed rather to opposite poles) is much larger than the
stellar radius. The same conclusion would follow from an intuitive
analysis of Fig.~\ref{3dd}, where it is seen that the multipolar
poles of an 'appropriate' polarity tend to bend the field lines in
their direction, thus to increase their curvature. On the
contrary, the opposite poles 'repel' and thus straighten the field
lines directed to them.
\begin{figure}
  \unitlength1.0cm
  \begin{center}
    \begin{picture}(17.,12.3)
      \put(-0.1,-0.6){
        \epsfysize=13cm
        \epsffile{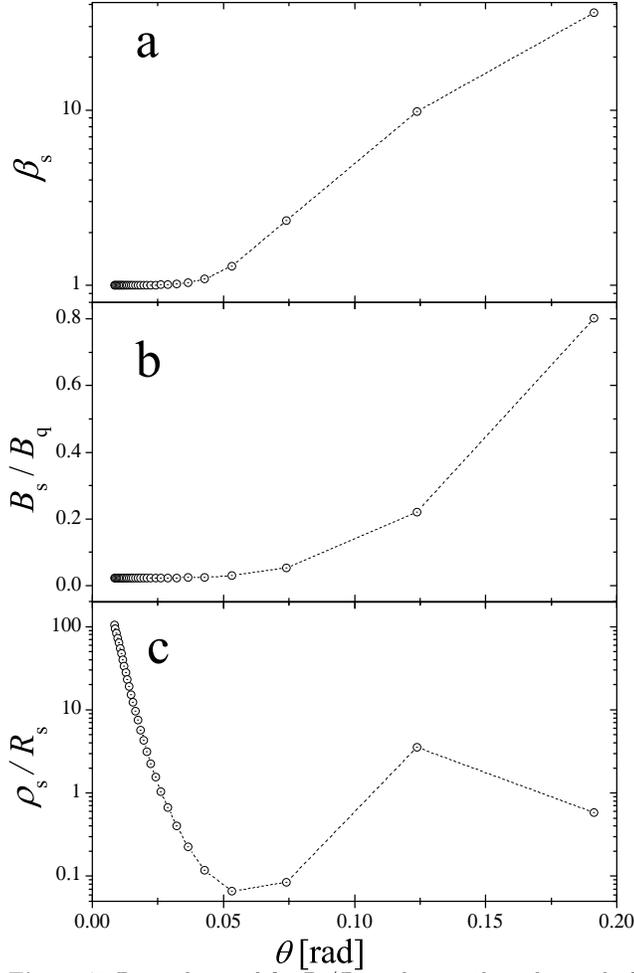}}
    \end{picture}
  \end{center}
\caption{Dependence of $\beta_{\mathrm s}$,
  $B_{\mathrm s}/B_{\mathrm q}$ and $\rho_{\mathrm s}$ on the polar
  angle $\theta$, in the direction of the local multipolar pole of
  order $l=4,m=4$.}
 \label{all_theta_4}
\end{figure}

Dependence of $\beta_{\mathrm s}$, $B_{\mathrm s}/B_{\mathrm q}$ and
$\rho_{\mathrm s}$ on the polar angle $\theta$ in the direction of an
'appropriate' multipolar pole is displayed in
Figs.~\ref{all_theta_4}(a), (b) and (c), respectively. These figures
show that $\beta_{\mathrm s}\geq 1$ and $\rho_{\mathrm s}\sim (0.1 \div
1) R_{\mathrm s}$ at the peripheral fraction of the polar cap. At the
same time, as we have seen in Figs.~\ref{all_phi_4}(a), (b) and (c),
the polar cap periphery is also modulated azimuthally.

Therefore, it can be concluded that the conditions for a copious
pair creation process are only fulfilled in localized regions,
symmetrically distributed around the dipolar magnetic axis, within
the thin peripheral part of the modified polar cap. Although at
high altitudes above the stellar surface the field lines strictly
follow the dipolar geometry, which is azimuthally uniform, such a
non-uniform pattern of plasma creation near the stellar surface is
to be reflected in the non-uniformity of the secondary plasma
density distribution within the cross-section of the open magnetic
flux tube at some large altitude ($r\geq r_{_{\mathrm i}}$). First
of all, the pair plasma will only flow along the outermost narrow
conical layer of the open dipolar flux tube. The radial
$\theta$-width of this layer can be estimated from
Fig.~\ref{all_theta_4}(c). Indeed, out of the 30 footprints of the
open field lines displayed in this figure, only outermost 10 of
them are feet of the field lines with $\rho_{\mathrm
  s}\sim (0.1 \div 1) R_{\mathrm s}$. Recalling that these points have
equal spacing at large altitudes and also taking into account the
boundary conditions~(\ref{init_cond_4}), one can estimate that the
width of the conical layer loaded by the pair plasma will
constitute about $(1-0.8)/3\approx 0.07$ fraction of the total
angular radius of the open flux tube. Moreover, this flow will
also be fragmented azimuthally: pair plasma will flow only along
the bundles of field lines which were stretched towards the local
multipolar poles while emerging from the stellar surface. In the
case of $l=4, m=4$, presented above, this corresponds to four
isolated bundles of field lines. Therefore, plasma flow will be
confined into a thin 'patchy hollow cone'. On the other hand, it
is well known that physical processes leading to the generation of
pulsar radio emission are associated with the existence of a dense
pair plasma. That being so, the resulting emission cone will
appear as split into symmetrically distributed 'hot spots'. This
is how, in our opinion, the 'memory' of the surface field
structure is preserved in the pulsar radio emission beam.

Let us notice that such a geometry is able to provide azimuthally
localized emission beams in both the vacuum gap model (Ruderman \&
Sutherland 1975) and its competing space charge limited flow model
(Arons \& Scharlemann 1979, modified by Muslimov \& Tsygan 1992).
As well-known, the latter model has previously been criticized
mainly because of its inability to provide isolated plasma
filaments and associated beams of emission. The inclusion of
multipolar magnetic fields with $m\neq 0$ removes the necessity of
introducing spark discharges above the polar cap, as complex
surface fields directly fragment the flow into isolated plasma
columns.

\subsection{Complex configurations containing
axially symmetric and non-uniform multipolar magnetic fields: a
model for PSR B$0943+10$} \label{PSR} A study of field line
topology and determination of curvature radii in a configuration
involving a multipole of high-order is proposed with intent to
interpret the special radio emission features observed for PSR
B$0943+10$, whose dynamical parameters are: $P=1.0977$~s and $\dot
P=3.529\times 10^{-15}$~s/s. We assume that, for some reason, the
open magnetic flux tube near the surface of this pulsar is only
dominated by the global dipolar field plus the multipolar magnetic
field with $l=20$ and $m=20$, an assumption which agrees with the
results of Mitra et~al. (1999).

The calculations according to the same scheme as in the previous
case were done for the field lines which, at the initial height
$r_{_{\mathrm i}}= 1.5 R_{\mathrm s}$ (where $B_{20,20}\ll B_d$),
constitute the outermost 0.1 fraction of the open field lines
tube. In other words, the boundary conditions were taken within
the following range:
\begin{eqnarray}
&&0.9\theta_{_{\mathrm i}}\leq \theta
(r_{_{\mathrm i}})\leq \theta_{_{\mathrm i}},\\ \nonumber
&&0\leq \phi (r_{_{\mathrm i}})\leq 2 \pi.
\label{init_cond_20}
\end{eqnarray}
Let us notice that, just like in the previous case, angular spacings
between the initial points within these ranges at the distance
$r_{_{\mathrm i}}$ were taken equal, both in azimuth $\phi$ and in
polar angle $\theta$.

\begin{figure}
  \unitlength1.0cm
  \begin{center}
    \begin{picture}(17.,5.1)
      \put(-0.1,-0.6){
        \epsfysize=5.5cm
        \epsffile{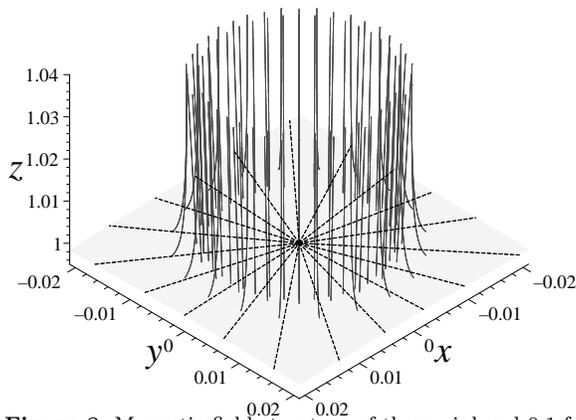}}
    \end{picture}
  \end{center}
\caption{Magnetic field structure of the peripheral 0.1
    fraction of the open flux tube near the surface of PSR B$0943+10$,
   assuming that the magnetic field is composed of the dipolar field
    and the multipole of order $l=20$ and $m=20$.}
 \label{3dd_20}
\end{figure}

The magnetic field structure of such a configuration near the
stellar surface is presented in Fig.~\ref{3dd_20}. Similarly to
the previous case, we observe that some field lines deviate
outwards, that is, towards the 'appropriate' local multipolar
poles, whereas other field lines deviate inwards. For clearness,
we again do not represent in this plot (and the other figures
below) the inner open field lines, which are less affected by the
multipolar field, and hence follow the dipolar geometry.
\begin{figure}
  \unitlength1.0cm
  \begin{center}
    \begin{picture}(17.,7.3)
      \put(-0.1,-0.6){
        \epsfysize=8cm
        \epsffile{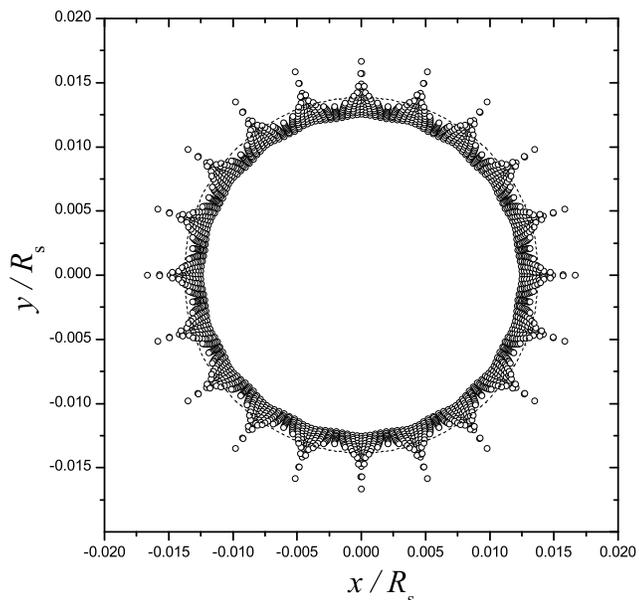}}
    \end{picture}
  \end{center}
\caption{Peripheral fraction of the modified polar cap of
    PSR B$0943+10$ assuming the contribution of the multipolar component
    with $l=20,m=20$. The footprints of the open field lines are
   represented by circles. The dashed circle shows, for comparison, the
    margin of the corresponding purely dipolar polar cap.}
 \label{polcap_20}
\end{figure}
\begin{figure}
  \unitlength1.0cm
  \begin{center}
    \begin{picture}(17.,7.3)
      \put(-0.1,-0.6){
        \epsfysize=8cm
        \epsffile{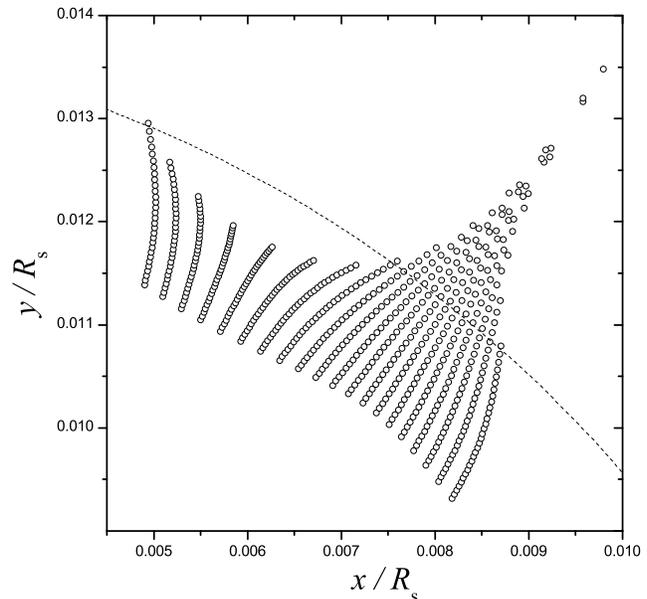}}
    \end{picture}
  \end{center}
\caption{A fraction of the modified polar cap (see
    Fig.~\ref{polcap_20}), representing one of the localized sub-regions
    where the curvature radius of the open field lines varies from
    'favourable' to 'unfavourable' values for the pair creation process.}
 \label{polcap_20_frac}
\end{figure}
\begin{figure}
  \unitlength1.0cm
  \begin{center}
    \begin{picture}(17.,12.3)
      \put(-0.4,-0.6){
        \epsfysize=13cm
        \epsffile{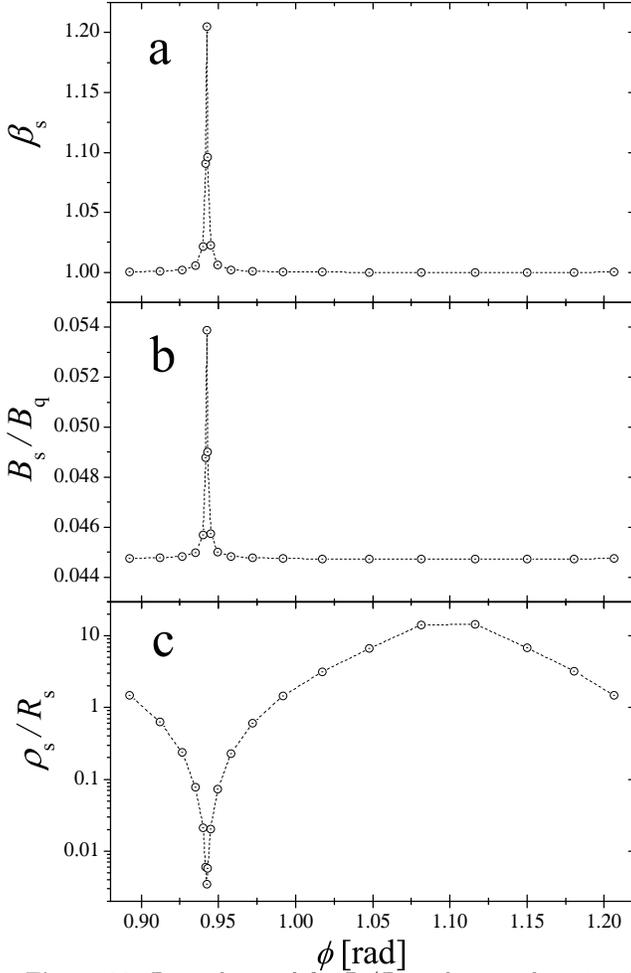}}
    \end{picture}
  \end{center}
\caption{Dependence of $\beta_{\mathrm s}$,
  $B_{\mathrm s}/B_{\mathrm q}$ and $\rho_{\mathrm s}$ on the
  azimuthal angle $\phi$, at the feet of the last open field lines of
  PSR B$0943+10$, assuming contribution of the multipole $l=20,m=20$.}
 \label{all_phi_20_frac}
\end{figure}
\begin{figure}
  \unitlength1.0cm
  \begin{center}
    \begin{picture}(17.,12.3)
      \put(-0.4,-0.6){
        \epsfysize=13cm
        \epsffile{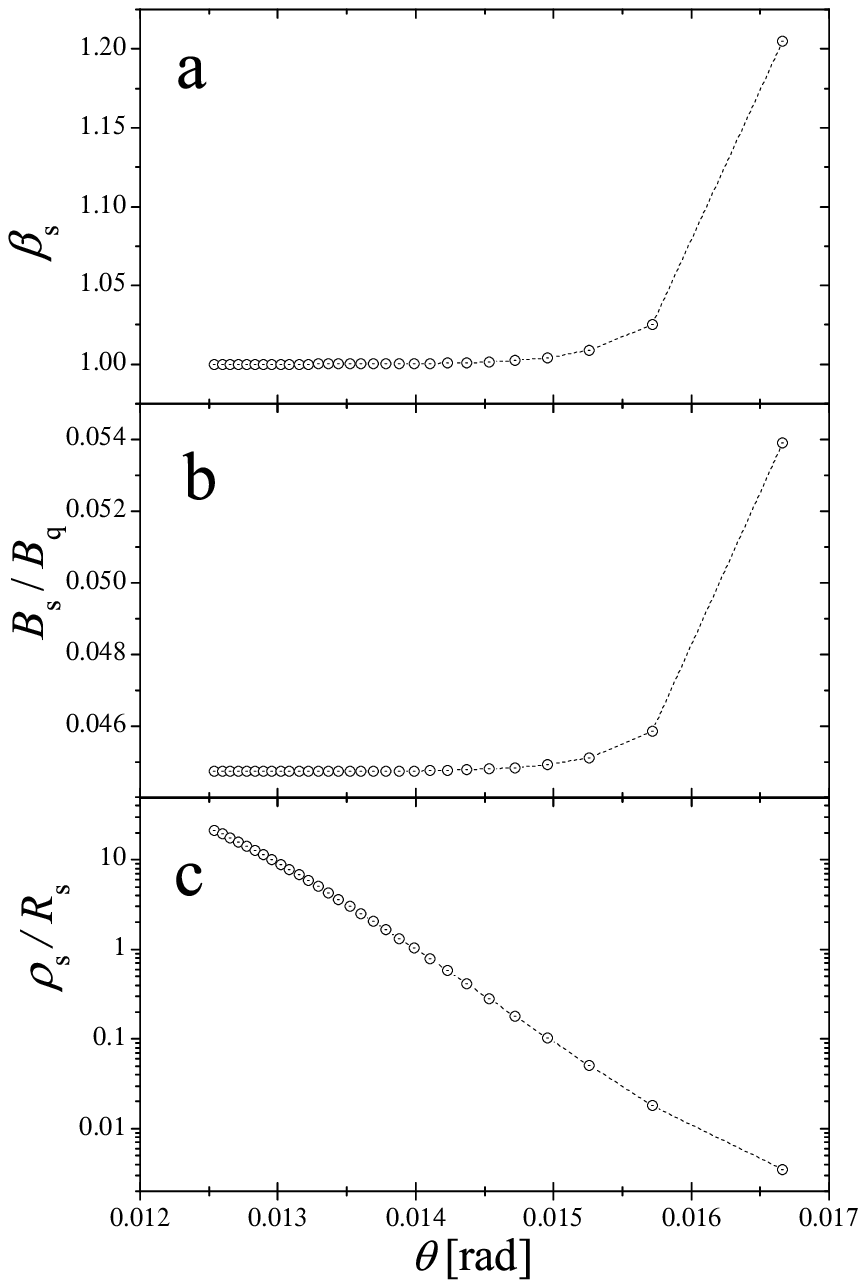}}
    \end{picture}
  \end{center}
\caption{Dependence of $\beta_{\mathrm s}$,
  $B_{\mathrm s}/B_{\mathrm q}$ and $\rho_{\mathrm s}$ on the polar
  angle $\theta$, in the direction of the local multipolar pole, for
  the same pulsar.}
 \label{all_theta_20}
\end{figure}

Fig.~\ref{polcap_20} represents the peripheral part of the
modified polar cap of this pulsar. It is clearly seen that the
outer margin of the polar cap appears to be fragmented into 20
symmetrically arranged sub-regions. One such particular sub-region
is plotted in detail in Fig.~\ref{polcap_20_frac}. The azimuthal
variation of the total magnetic field strength measured in the
units of, first, the dipolar field, and then, the critical field,
as well as the variation of the curvature radius of the last open
field lines within this sub-region are displayed in
Figs.~\ref{all_phi_20_frac}(a), (b) and (c), respectively. The
variation of $\beta_{\mathrm s}$, $B_{\mathrm s}/B_{\mathrm q}$
and $\rho_{\mathrm s}$ with the polar angle $\theta$ towards an
'appropriate' multipolar pole within this sub-region is presented
in Figs.~\ref{all_theta_20}(a), (b) and (c), respectively. It is
seen in these figures that the conditions for the development of
pair cascades are fulfilled only above the restricted areas of the
polar cap periphery.

At large altitudes from the stellar surface, the 'memory' of the
surface field structure will be preserved in the form of localized
plasma columns distributed symmetrically around the outer margin
of the open magnetic flux tube. Similarly to the previous case of
$l=4$ and $m=4$, the radial width of this thin outer ring can be
estimated from Fig.~\ref{all_theta_20}(c), using the boundary
conditions~(\ref{init_cond_20}): it will constitute about 1/20
fraction of the total radius of the open field lines tube. This
'hollow cone' will also be fragmented azimuthally, and the
transverse dimension of an individual plasma column can be
estimated from Fig.~\ref{all_phi_20_frac}(c) as being about $
1/40$ fraction of the circumference. These plasma columns will be
separated by low-density 'slits' of approximately the same
transverse angular dimension.

Obviously, such a structure of the open flux tube exhibits a
striking similarity with the 20-fold pattern of emission sub-beams
discovered by Deshpande \& Rankin (1999, 2001) using their
'cartographic transformation' technique. They found an identical
configuration at two different frequencies, which suggests a
filamentary plasma distribution rather than an emission process,
as a possible origin for such a pattern. Deshpande \& Rankin
(2001) conclude that such plasma columns should have feet within a
certain ring on the polar cap. In view of our multipolar model,
such a ring could be associated with the outer margin of the
modified polar cap, represented in Fig.~\ref{polcap_20}.

According to Deshpande \& Rankin (1999), 20 emission sub-beams undergo
rigid counter-clockwise rotation around the magnetic axis in a total
time of $37 P = 41$~s, maintaining their number and spacing despite
perturbations tending both to bifurcate a given beam and to merge
adjacent ones. Although at this stage it is only possible to speculate
about several possible reasons for this effect, we suggest in the next
section an explanation of such a rotation, in terms of the character
of the perturbations in the azimuthal direction.

In another way, such a phenomenon could actually reflect some
stroboscopic-like effect. Indeed, pair creation in one of the
sub-regions in Fig.~\ref{polcap_20} results in a roughly Gaussian
distribution of the pair plasma density across the local magnetic
field. Such a distribution function would screen the electric
field, at least partially, in the adjacent sub-regions, thus
temporarily ceasing or, at least, largely reducing the pair
production process there. The situation would revert after pair
plasma cascade will exhaust in this sub-region, when the
accelerating electric field is screened out. Then, two adjacent
sub-regions will 'turn on', reducing pair creation in their
adjacent sub-regions. Such consecutive 'switches', in combination
with the $\bmath{E\times B}$-drift and the polar cap heating, may
result in the above-mentioned stroboscopic effect.

It should be noticed that, in order to obtain the patterns
presented in Figs.~\ref{polcap_20} $-$ \ref{all_theta_20}, one has
to choose the free parameter $\xi$ (introduced in
equation~\ref{B_lm_new}) in our calculations so that the magnetic
field strength in the close vicinity of the multipolar poles
themselves appears to exceed significantly the critical field
$B_{\mathrm q}$. However, these multipolar poles are located at
lower latitudes of the stellar surface, close to the neutron
star's equator. Therefore, very high magnetic field is achieved
well within the domain of the closed field lines where plasma
creation processes certainly do not occur. Besides, there are no
serious theoretical limitations to the strength of the magnetic
field in nature (Usov V. V., private communication).

\section{A model for radio emission from an ensemble of finite sub-beams}
\label{Model}
\subsection{Features of the model}
\label{Model-Features} As demonstrated in Section~\ref{Topology:
m/=0}, the particular magnetic field topology which results from
the superposition of the global dipolar magnetic field and a
high-order multipolar field with $m\neq 0$, both with high
strength, suggests that the pair plasma created at the bases of
magnetic flux tubes is guided within isolated thin filaments,
located at the outer margin of the bundle of open magnetic field
lines. According to Sturrock (1971), the density of the pair
plasma within the flux tubes is proportional to the Goldreich \&
Julian (1969) density, namely $n_{\mathrm p} = \kappa n_{_{\mathrm
GJ}}$, the $\kappa $ parameter being in the domain $10^{3} \leq
\kappa \leq 10^{5}$. Outside, the plasma is supposed to have a
much smaller density, of the order of the Goldreich \& Julian
density, $n_{_{\mathrm GJ}}$. Thus, the whole system of dense
filaments can be described as an ensemble of relativistic finite
beams immersed in external low-density media. These external media
in the pulsar magnetosphere are, on the one hand, the charged
static medium lying on closed field lines, and on the other hand,
the primary beam formed of ultra-relativistic particles of one
charge flowing inside the hollow cone and between the dense beams.

The simplest case concerns one flowing annular beam of
relativistic pair plasma surrounded by external plasmas.  The
dispersion relation giving the characteristics of perturbations
able to propagate in such a system were obtained matching the
solutions for the electromagnetic field components at the
different beam-plasma interfaces (Asseo 1995). The case of the
ensemble of twenty isolated sub-beams discussed in
Section~\ref{PSR} is more complex: the beams are supposed to be
regularly distributed over a ring that delimits the emission
region; they have different radial and azimuthal extents and are
separated over the ring by distances of the order of their
azimuthal extent. The radial variation of the density operates in
two different ways: either there is an abrupt change of the
density at the edges of the ring, or there is none. The azimuthal
changes of the density from one beam to the next one are alike.
The geometrical characteristics of the relativistic beams are
evaluated taking into account the fact that the observed ensemble
of twenty beams have their origin in the gap region, very close to
the stellar surface, where the complex multipolar field determines
the pair creation process. As concluded in the previous section
(see Figs.~\ref{all_theta_20}(c) and \ref{all_phi_20_frac}(c) for
details), the radial angular width of an individual sub-beam
constitutes a fraction equal to about 1/20 of the total radius of
the open field lines tube at an altitude $r_{_{\mathrm i}}$ (where
$B^{lm}\ll B^{\mathrm d}$), whereas its azimuthal angular width
represents a fraction equal to about 1/40 of the circumference.
Therefore, at some distance $r_{_{\mathrm i}}$ (see
Section~\ref{Topology: general}), the radial and azimuthal extents
of individual sub-beams can be estimated as, respectively,
\begin{equation}
w_{\theta}= \frac{r_{_{\mathrm i}}\sin\theta_{_{\mathrm i}}}{20} \approx
7.3\times 10^{2} P^{-1/2} \left(\frac{r_{_{\mathrm i}}}
{R_{\mathrm s}}\right)^{3/2}\ \mathrm{[cm]}
\label{w_r}
\end{equation}
and
\begin{equation}
w_{\phi}= \frac{2 \pi r_{_{\mathrm i}}\sin\theta_{_{\mathrm
i}}}{40} \approx 2.3\times 10^{3} P^{-1/2}
\left(\frac{r_{_{\mathrm i}}} {R_{\mathrm s}}\right)^{3/2}\
\mathrm{[cm]}, \label{w_phi}
\end{equation}
where equation~(\ref{theta_max}) has been used in order to
evaluate $\theta_{_{\mathrm i}}$.  For example, at a distance
$r_{_{\mathrm i}} = 2 R_{\mathrm s}$, far above the gap region,
the numerical estimates for the individual sub-beam width yield
$w_{\theta}\approx 2\times 10^{3}$~cm and $w_{\phi} \approx
6.5\times 10^{3}$~cm, respectively, for a pulsar with period $P =
1$~s.  The ratio of the azimuthal and radial sub-beam widths is
constant, namely $w_{\phi}/w_{\theta} \approx \pi$ whatever the
distance, but the typical wavelengths associated with observed
radio frequencies are smaller than these characteristic widths.
Thus, one should test whether it is sufficient to consider
two-dimensional perturbations of the whole system of beams
immersed in exterior media, or if it is necessary to analyse the
fate of 3-dimensional perturbations.  This can be checked as in
Asseo, Pellat \& Sol (1983), considering the strongest
instabilities which may develop in a thin beam surrounded by
different plasmas.

In the previous sections we used the spherical system of coordinates
$(r, \theta,\phi)$, centered at the stellar centre. Instead, in this
section it is convenient to use a cylindrical system of coordinates
$(\varrho, \varphi, z)$, adequate for a local description. Here the
radial coordinate $\varrho$ is measured along the local curvature
radius of the magnetic field line; $\varphi$ is the azimuthal angle
which corresponds to the angular variation along the circle with the
radius $\rho$, this circle being tangent to the magnetic field line
passing through the point $(r, \theta)$ in the plane $\phi=\phi_0$ of
the spherical geometry; $z$ is the coordinate perpendicular to the
plane $\phi =\phi_0$.

Assuming 3-dimensional perturbations in the cylindrical system
depending on ($\varrho, \varphi, z$) and characterized by
wave-numbers $m$ and $k_{z}$, the strongest instabilities are
obtained at the resonant frequency $\omega \approx m c/ \rho$ and
for transverse wave-numbers in the domain $ \Delta k_z \leq
\sqrt{2} m^{2/3} \rho^{-1}$. In this case the analytical solutions
obtained for the electromagnetic fields characterizing the
perturbation in terms of Bessel and Hankel functions, result in an
unmodified growth rate of the instabilities. Consequently, they
are valid for beams with a spatial extent
\begin{equation}
\Delta z = w_{\phi} \geq \Delta z_{_{\mathrm min}}\equiv
\sqrt{2}\pi m^{-2/3} \rho,
\label{fin}
\end{equation}
where we used the fact that the spatial extent of a sub-beam $\Delta
z=2\pi/\Delta k_z$ is nothing else but its azimuthal width $w_{\phi}$
(equation~\ref{w_phi}).

\subsection{The case of millisecond and fast pulsars}
\label{Model-msp}
\begin{figure}
  \unitlength1.0cm
  \begin{center}
    \begin{picture}(17.,12.3)
      \put(-0.1,-0.6){
        \epsfysize=13cm
        \epsffile{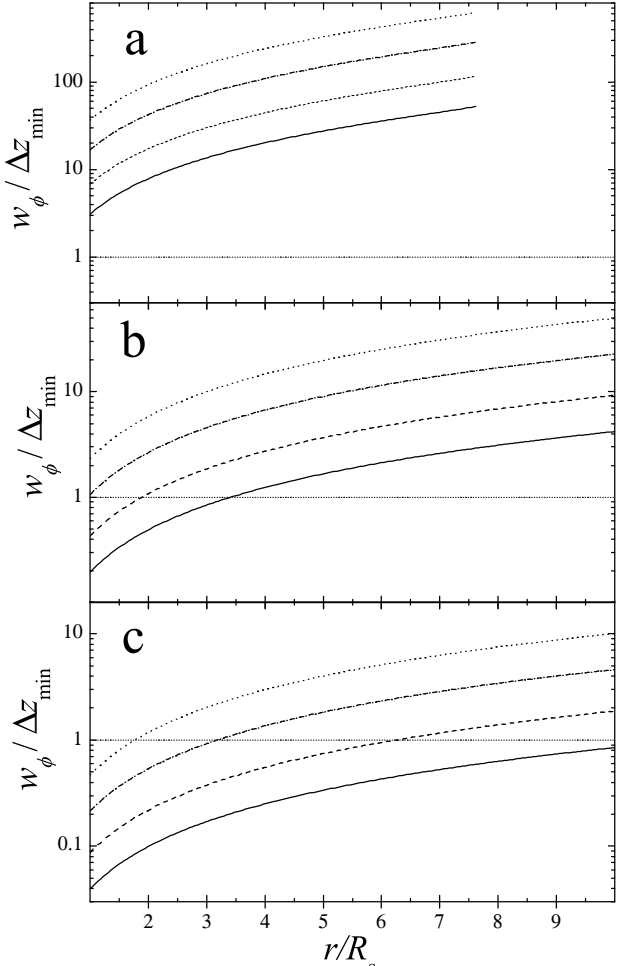}}
    \end{picture}
  \end{center}
\caption{The condition~(\ref{fin}) at different
  distances from the stellar centre. The panels (a), (b) and (c)
  represent the results of calculations for the sample pulsars with
  periods equal to $1.6\times 10^{-3}$~s, $0.1$~s and $1.0977$~s (PSR
  B$0943+10$), respectively. In each of the panels, solid curves
  correspond to the observed frequency $\nu=34$~MHz,
  dashed curves to $\nu=111.5$~MHz, dashed-dotted
  curves to $\nu=430$~MHz and the dotted curves to
  $\nu=1.4$~GHz.}
 \label{finite}
\end{figure}

Numerical estimates show that in the case of
millisecond and fast pulsars the constraint~(\ref{fin}) is easily
fulfilled just above the gap region and beyond, considering the
domain of observed radio frequencies $\nu=2\pi/\omega$ (see
Figs.~\ref{finite}(a) and (b), respectively). This indicates that
in such a case, finite transverse dimensions do not modify the
results obtained in cylindrical geometry using two-dimensional
perturbations to characterize radio emission features generated
above the gap region within an annular hollow conical beam, or
within an ensemble of discrete sub-beams similar to those observed
in PSR B$0943+10$.

In such conditions, one may consider that locally the propagating wave
only sees the radial variations of the density, so that it is possible
to treat the problem in a poloidal plane. While doing so, we ignore
the density variations in the direction corresponding to the azimuthal
$\phi$-direction of the spherical geometry, that is to say, in the
transverse $z$-direction of the cylindrical geometry, and treat each
beam of the ensemble of beams as isolated and immersed in 'infinite'
exterior media. This allows us to extend the results which concern the
characteristics of the perturbations able to propagate in the system
formed by relativistic beam and plasma flows which fill a hollow
conical region and are immersed in exterior media, to the more complex
situation of the ensemble of beams.

The dispersion relation specific to the case of a finite beam
immersed in external media was obtained by matching the solutions
for the electromagnetic field components at the different
beam-plasma interfaces (Asseo 1995).  It can be expressed in terms
of the dielectric constant for the relativistic beam plus
relativistic plasma flows, $W$,
and of the dielectric constants of the exterior media,
$W_{i}$ and $ W_{e}$.
It is obtained close to the resonant frequency,
\begin{equation}
\omega \approx m \Omega_{0} + \delta \omega \approx
\omega_{_{\mathrm R}}  + \delta \omega   \approx
\omega_{_{\mathrm R}}(1 + Y),
\end{equation}
where $m \Omega_{0} \approx m c /\rho $ is the frequency associated
with the circulation of relativistic particles along magnetic flux
tubes and $Y = {\delta \omega}/{\omega_{_{\mathrm R}}} \leq 1$ is the
relative growth rate. Assuming variations with the radial distance
together with a wave-like behaviour proportional to $\exp{ [-{\mathrm
    i}(\omega t - m \phi)]}$ for linear perturbations of the whole
system, the dispersion relation becomes:
\begin{eqnarray}
\tanh \left\{\frac{2 m W^{1/2}}{3} \left[\left(-Y \right)^{3/2} -
\left(\frac{w_{\theta}}{\rho} - Y\right)^{3/2}\right]\right\}
= \nonumber \\
W^{1/2} \frac{W_{i}^{1/2}  +  W_{e}^{1/2}}{W + W_{i}^{1/2}
W_{e}^{1/2}}.
\label{dispersion}
\end{eqnarray}
Let us notice that the radial width in the cylindrical system of
coordinates, $w_{\varrho}$ is approximately equal to the angular
width $w_{\theta}$ in the spherical geometry, so that the above
dispersion relation is correct in a cylindrical geometry.  Such a
dispersion relation for finite beam and plasma flows differs from
the dispersion relation that characterizes the two-stream
instability obtained for infinite and homogeneous beam and plasmas
described in a straight geometry, namely $W =0$. It is modified
due to the presence of external media with dielectric constants
$W_{i}$ and $W_{e}$. It is also modified due to geometric
properties of the flow: more precisely the curvature of the
magnetic field, as the radius of curvature $\rho$ of magnetic
field lines is involved, the finite width of the beam $w_{\theta}$
and the azimuthal wave-number of the perturbation $m$. The
analysis of this modified dispersion relation in different limits
shows that it allows the description of both the radiative (see
Goldreich \& Keeley (1971) who first described it) and two-stream
instabilities, well-known in the context of pulsar physics, but
also that there is an unexpected instability: the 'finite beam
instability', which even develops for a one-component beam of
relativistic mono-energetic particles bounded by external media.

The characteristics of the 'finite beam instability' were obtained
in the most general conditions for a beam bounded by identical, or
non-identical, external media from the above dispersion relation
(Asseo 1995).
%
%
%
%
%
The case of a beam bounded by identical external media with $W_{i}
= W_{e}$, is of interest here for an ensemble of beams delineated
by the superposition of a strong multipole magnetic field plus a
dipole field.
In the 'thin' beam case with $Y \gg w_{\theta}/ \rho$, the
dispersion relation simplifies to:
\begin{equation}
\left(\frac{W}{W_{i}}\right)^{1/2}
\tanh \left(\frac{1}{{\mathrm i}\epsilon\sqrt{2}}W^{1/2}
  Y^{1/2}\right) =  1.
\label{dispersion_thin}
\end{equation}
The dependence on the presence of external plasmas appears through
$W_{i}$. The dependence on the geometric properties of the flow
appears through the parameter:
\begin{equation}
\epsilon  = \frac{\rho}{ m } \frac{1}{ w_{\theta}}.
\end{equation}

%
%

For the magnetic configuration which includes both dipolar and
multipolar magnetic field components, the value of the radius of
curvature of magnetic field lines varies with the distance from the
stellar surface.  Within the gap region from which the beams are
issued,
the radius of
curvature of magnetic field lines is of the order, or even less, than
the radius of the star, $\rho \leq R_{\mathrm s}$. Beyond,
in the region where the dipolar magnetic field is dominant, the
radius of curvature is approximately $\rho\sim 10^{2} R_{\mathrm
s}$, (indeed, $\rho \approx (4 r/3\theta) \approx  10^{8}$~cm at
the distance $r= 2 R_{\mathrm s}$). Therefore, the azimuthal
wavenumber which corresponds to the resonant frequency at which
the instabilities are triggered off, fixed by the relation $m
\approx {\rho}{\omega_{_{\mathrm R}}/c}$, slightly varies in the
radio emission domain which starts above the gap region (see Asseo
1995). At such resonant frequencies,
\begin{equation}
\epsilon  = \frac{\rho}{m} \frac{1}{w_{\theta}}
 = \frac{c}{\omega_{_{\mathrm R}}} \frac{1}{w_{\theta}}
\end{equation}
only depends on the radial extent of the involved annular beam or
sub-beams.
Numerical estimates with usual parameters for a pulsar with
$P=1$~s show that in the domain of observed radio frequencies
$\omega_{_{\mathrm R}} \leq 1.6\times 10^{9}$~Hz and for the
radius of curvature of magnetic field lines in the domain $\rho
\leq 10^{8}$~cm, $\epsilon$ is smaller than unity.
%
%

Above the gap region, but relatively close to it, the ratio of the
ring width to the radius of curvature of magnetic field lines is
very small, $w_{\theta}/\rho \leq 2\times 10^{-5}$, for values of
the radius of curvature in the domain $\rho \leq 10^{8}$~cm. Thus,
we are in the thin beam case, the dispersion
relation~(\ref{dispersion}) is valid in its limiting form
(equation~\ref{dispersion_thin}),
and the approximation $\tanh \{\ldots\}\approx 1$ is adequate, the
parameter $\epsilon$ being much smaller than unity. In this case, the
dispersion relation~(\ref{dispersion}) simplifies to $ W = - W_{i}$.
Such a dispersion relation is adequate to describe a limited
region of emission whatever its altitude.

For a one-component beam of relativistic particles with Lorentz
factors $\gamma_{p}$, the dielectric constant $W$ is defined as:
\begin{equation}
W  = 1 - \frac{{\mathcal P}^{2}(r)}{Y^2},
\end{equation}
and the dielectric constant $W_{i} $ of the interior plasma supposed
to be at rest is defined as,
\begin{equation}
W_{i} = 1 - {\mathcal P}_{i}^{2}(r)(1 - 2 Y).
\end{equation}
Here,
\begin{equation}
{\mathcal P}(r) = \frac{\omega_{p}(r)}{\gamma_{p}^{3/2} m
\Omega_{0}},\ \ {\mathcal P}_{i}(r) = \frac{\omega_{pi}(r)}{m
\Omega_{0}}, \label{Pr}
\end{equation}
whereas $\omega_{p}(r) = \left[{4 \pi n_{p}(r)
    e^{2}}/{m}\right]^{1/2}$ and $\omega_{pi}(r)= \left[{4 \pi
    n_{pi}(r) e^{2}}/{m}\right]^{1/2}$, the plasma frequencies in the
flowing beam and in the exterior media, respectively, vary along
magnetic flux tubes with the corresponding densities.
Thus, ${\mathcal P}(r)$ and ${\mathcal P}_{i}(r)$ depend on the
densities in the interior and exterior flux tubes, respectively.
Assuming that, as mentioned above, the density within the magnetic
flux tube is proportional to the Goldreich \& Julian (1969) density,
\begin{equation}
n_{p}(r) =\kappa n_{_{\mathrm GJ}}(r),
\end{equation}
whereas the density in the exterior media $n_{pi}(r)$ is equal to the
Goldreich \& Julian density,
\begin{equation}
n_{_{\mathrm GJ}}(r) = \frac{B_{\mathrm s}}{e c P}
\left(\frac{R_{\mathrm s}}{r}\right)^3 \approx 7\times
10^{-2}\frac{B_{\mathrm s}}{P} \left(\frac{R_{\mathrm
s}}{r}\right)^3\ {\mathrm [cm]}^{-3},
\end{equation}
we have,
 \begin{equation}
{\mathcal P}(r) \approx 1.5\times 10^4 \sqrt{\kappa\frac{B_{\mathrm s}}{P}
\left(\frac{R_{\mathrm s}}{r}\right)^3}
\frac{1}{\gamma_{p}^{3/2}  m \Omega_0},
 \end{equation}
 \begin{equation}
{\mathcal P}_i(r) \approx 1.5\times 10^4\sqrt{\frac{B_{\mathrm s}}{P}
\left(\frac{R_{\mathrm s}}{r}\right)^3}\frac{1}{m \Omega_0}.
\end{equation}

Let us note that $W_{i}$ varies according to the value of
${\mathcal P}_{i}(r)$ in the emission region, namely:
\begin{enumerate}
\item $ W_{i} \approx 2 Y$ at the transition zone located at the
  distance $r \approx r_{_{\mathrm lim}}$ where ${\mathcal P}_{i}(r)
  \approx 1$.
\item $W_{i} \approx - {\mathcal P}_{i}^{2}(r)$ close to the surface
  of the star where ${\mathcal P}_{i}(r) \gg 1$, at distances $r <
  r_{_{\mathrm lim}}$.
\item $W_{i}\approx 1$ beyond the transition zone where ${\mathcal
    P}_{i}(r) \ll 1$, at distances $r > r_{_{\mathrm lim}}$.
\end{enumerate}

Including a multipolar magnetic field superimposed over the dipolar
field and assuming that at the surface of the star the intensity of
the multipolar magnetic field is proportional to the strength of the
dipole field, $B_{\mathrm s}^{lm} = (\beta_{\mathrm s}-1) B_{\mathrm
  s}^{\mathrm d}$ (where $\beta_s$ has been defined in
equation~\ref{beta}), we have for the total field, $B(r) =
B_{\mathrm s}^{\mathrm d}(R_{\mathrm
  s}/r)^{3} + (\beta_{\mathrm s} -1) B_{\mathrm s}^{\mathrm d}(R_{\mathrm
  s}/r)^{l+2}$. Therefore, the distance $r_{_{\mathrm lim}}$, at which
${\mathcal P}_{i}(r)\approx 1$, is not significantly different
from the distance $r_{_{\mathrm lim}}$ obtained in the case where
there are no multipolar field components in addition to the
dipolar field, as the relative contribution of the multipolar
magnetic field, $(\beta_{\mathrm s}-1) (R_{\mathrm s}/r)^{l-1}$,
is very small. In effect, $r_{_{\mathrm lim}}$ is simply
determined from the relation
\begin{equation}
{\mathcal P}_{i}(r_{_{\mathrm lim}}) = x f\left(r_{_{\mathrm lim}}\right)
= 1,
\label{rlim}
\end{equation}
where
\begin{equation}
x = 1.5\times 10^4\sqrt{\frac{B_{\mathrm s}^{\mathrm d}}{P}}\frac{1}{m
\Omega_{0}}
\label{xlim}
\end{equation}
and
\begin{equation}
f(r) = \sqrt{\left[ 1+\left(\beta_{\mathrm s}-1 \right)
\left(\frac{R_{\mathrm s}}{r} \right)^{l-1}\right]
\left(\frac{R_{\mathrm s}}{r} \right)^3}.
\label{flim}
\end{equation}
$r_{_{\mathrm lim}}$ increases with the strength of the total magnetic
field but decreases as the pulsar period and the frequency associated
with the circulation of relativistic beam particles increase.
%

In the extreme case of a millisecond pulsar with parameters $P=
0.0016$~s, $B_{\mathrm s}^{\mathrm d} = 10^8$~G, $m \Omega_0 =
10^9$~Hz, we obtain $r_{_{\mathrm lim}} = 2.4\times 10^6$~cm. For a
'normal' pulsar with parameters $P= 0.1$~s, $B = 10^{11}$~G, $m
\Omega_0 = 10^{9}$~Hz, we obtain $r_{_{\mathrm lim}} \approx 6\times
10^{6}$~cm.  For a 'slower' pulsar with parameters $P= 1$~s, $B =
10^{11}$~G, $m \Omega_0 = 10^{9}$~Hz, we obtain $r_{_{\mathrm lim}}
\approx 3\times 10^{6}$~cm. Therefore, whatever the type of pulsar,
$r_{_{\mathrm lim}} \approx$ a few $R_{\mathrm s}$.

The distance $r_{_{\mathrm lim}}$ is important in delineating the
different regions of instability: below $r_{_{\mathrm lim}}$ the
'finite beam instability' is dominant whereas beyond $r_{_{\mathrm
    lim}}$ either the radiative or the two-stream instabilities
prevail. As a matter of fact, the behaviour and growth rate of the
'finite beam instability' is different according to the distance
of the considered emission region to the stellar surface. The
characteristics of the unstable waves excited trough the 'finite
beam instability' -- frequency, polarisation, available
electromagnetic energy and growth rate -- are in agreement with
pulsar radio observations (Asseo 1995). Thus, the 'finite beam
instability' will initiate the radio emission process in the
ensemble of beams with their origin in the gap region, relatively
close to the surface of the star.  Moreover, such an instability
strongly depends on the ratio of densities at the basis of the
magnetic tube within and outside the limited beam. As the density
of the created pair plasma may vary, an observed apparent motion
of the beams around the magnetic axis could be related to
variations of the densities within and outside the successive
tubes.

\subsection{The case of slow pulsars and PSR B$0943+10$}
\label{Model-slow}
Numerical estimates concerning the above constraint on the spatial
extension of the beams, required to ignore azimuthal variations of the
perturbations (equation~\ref{fin}), show that in the case of PSR
B$0943+10$ this constraint is not uniformly satisfied in the whole
region above the gap (see Fig.~\ref{finite}(c)). The distance
$r_{_{\mathrm 2D}}$ from the center of the star beyond which the
constraint is satisfied varies with the frequency of observation: as
clear from Fig.~\ref{finite}(c), larger distances correspond to lower
frequencies. Indeed, $r_{_{\mathrm 2D}} \approx 2.3\times 10^{6}$~cm
for a frequency of observation $\nu=\omega/2\pi \approx 430$~MHz,
$r_{_{\mathrm 2D}} \approx 4.6\times 10^{6}$~cm for $\nu \approx
111.5$~MHz and $r_{_{\mathrm 2D}} \approx 8.7\times 10^{6}$~cm for
$\nu \approx 34$~MHz.

Consequently, between the gap region and the location $r =
r_{_{\mathrm 2D}}$, it is necessary to take into account the
3-dimensional character of the perturbations able to propagate in
the system of discrete sub-beams: the density variations in the
direction corresponding to the azimuthal $\phi$-direction of the
spherical geometry, that is to say, in the transverse
$z$-direction of the cylindrical geometry, cannot be further
ignored and the ensemble of sub-beams has to be considered as a
whole system immersed in 'infinite' exterior media.  Thus, an
additional systematic azimuthal variation of the perturbations
below $r_{_{\mathrm 2D}}$ will exist and modify the character of
perturbations beyond $r_{_{\mathrm 2D}}$, in the domain where
2-dimensional perturbations are dominant.  In fact a coupling of
the different perturbed physical quantities is involved as soon as
one introduces an azimuthal dependence of the perturbations at
relatively small distances from the surface of the star. A
thorough treatment including the dependence on the coordinates
$(r, \phi, z)$ of the specified cylindrical geometry of the
perturbations is possible as in Asseo et al. (1983), writing the
perturbations as of the type:
\begin{equation}
  g(z,r)
 \exp{ [-{\mathrm i} k_{z} z]}
  \exp{ [-{\mathrm i}(\omega t - m \phi)]}.
\end{equation}
Obviously, the character of the perturbations in the region
located beyond $r_{_{\mathrm 2D}}$ has to approach the one
described above for fast pulsars, whereas in the region below
$r_{_{\mathrm 2D}}$ it is modified by the dependence on the
variable $z $ and wave-number $k_{z}$. Although we here restrict
to a qualitative explanation, and leave exact calculations
concerning the character of 3-dimensional perturbations for future
work, this suggests that for PSR B$0943+10$, the observed
'rotation', and/or 'drifting', of the emitting columns may be a
trace of the initial waves formed in the region below
$r_{_{\mathrm 2D}}$ and of their transition towards the above
region.

Most of the pulsars which exhibit beautiful drifting subpulses --
PSRs $0809+74$,
$0320+39$,
$0820+02$,
$0818-13$,
$1540-06$ and
$0943+10$ -- show periodic drifts (as observed and/or classified
by Rankin (1986), see table~2 therein).
We notice that all the quoted pulsars have sufficiently long
periods to be modelled in analogy of PSR B$0943+10$. This suggests
that 'drifting' effects, already known as intrinsically conal
phenomena and associated with the circulation of disturbances
within the annular polar cap region, could also result from
similar perturbations moving within a hollow conical region
fragmented into discrete sub-beams.  Then, azimuthal widths of the
sub-beams will be reflected in the character of unstable excited
perturbations.

\section{Summary}\label{Summary}
In this paper we study the complex magnetic topology in the
vicinity of neutron stars. We restrict to the case of the aligned
rotator, assuming parallel, or anti-parallel, orientation of the
rotation and magnetic axes. We assume that the total magnetic
field results from the superposition of the global dipolar field
and of star-centered multipolar fields of comparable, or even of
higher strength. We obtain the following results:
\begin{enumerate}
\item For configurations involving axially symmetric and uniform
multipolar magnetic fields it is not possible to obtain small
curvature radii of the magnetic field lines, even for multipoles
of higher order and strength relative to the dipolar field. Let us
recall that according to pulsar polar gap models, curvature radii
of the order of the stellar radius are crucial for an efficient
pair creation process.
\item An interesting consequence of such a magnetic topology is the
  existence of 'neutral points' and 'neutral lines' above one of the
  pulsar polar caps. We plan to study implications of such peculiar
  features of the magnetic field geometry in future works.
\item For configurations involving axially symmetric and non-uniform
  multipolar magnetic fields, the magnetic field structure just
  above the polar cap is modified in such a way that the pair plasma flow at high
  altitudes appears as modulated in separate filaments regularly
  distributed around the outer margin of the open flux tube.  Such
  filaments have their feet in the modified polar cap.  They are thin
  and have finite radial and azimuthal extents. Therefore, the
  non-uniform pattern of the distribution of curvature radii over the
  polar cap is reflected in the inhomogeneous distribution of the pair
  plasma density within the cross-section of the open flux tube at
  high altitudes.
\item For the particular superposition of a dipolar field and a
  multipolar field of order $l= 20$ and $m=20$, the magnetic topology
  causes fragmentation of the pair plasma flow into twenty isolated
  thin filaments.  This resembles the set of twenty sub-beams observed
  by Deshpande \& Rankin (1999, 2001) in PSR B$0943+10$.
\item Emission in such a system of isolated thin beams, flowing along
  the curved magnetic field lines, can be described in terms of
  unstable waves excited by specific instabilities. Relativistic
  finite beams flowing in the pulsar magnetosphere along a limited
  fraction of the hollow cone surrounding the magnetic axis can be
  treated as immersed in exterior media, namely the medium in the
  closed field line region and the medium within the hollow cone of
  the open magnetosphere (Asseo 1995). Three different instabilities
  may develop in such finite beams, depending on the distance to the
  center of the star: the 'finite beam', radiative and two-stream
  instabilities. As previously obtained by Asseo (1995), close to the
  surface of the star the 'finite beam' instability will generate
  waves whose properties (frequency, polarisation and available
  electromagnetic energy) are in agreement with the observed pulsar
  radio emission.
\item We check that the conditions for such a treatment are relevant
  for the configuration of discrete sub-beams that may exist in
  millisecond and fast pulsars. They are also relevant for slower
  pulsars, like PSR B$0943+10,$ beyond some distance from the gap region.
  On the other hand, closer to the gap of such a slow pulsar, the radial and
  azimuthal variations of the flowing sub-beams cannot be separated;
  the system of sub-beams has to be treated as a whole; the
  3-dimensional character of the perturbations will introduce some
  'drift' of the emitted waves; such a 'drift' effect will be
  transferred to high altitudes due to the necessary continuity between the
  vicinity of the gap and the farther region. In this way, the
  continuity of the flow all along magnetic flux tubes and of the
  perturbations carried by it, accounts for the 'drifting' character of
  the subpulses observed in pulsar radio
  profiles.
\end{enumerate}

\section{Acknowledgments}
It is a pleasure to thank our referee for reading carefully our
manuscript and suggesting changes which have been fruitful to
improve our paper. D.K. would like to thank G. Melikidze, V. Usov
and J. Gil for fruitful discussions. D.K. is also very grateful to
the scientific staff of the Centre de Physique Th{\'e}orique (CNRS
UMR 7644) for hospitality during his stay at Ecole Polytechnique.

\end{document}